\documentclass[12pt]{article}

\makeatletter
\newcount\@tempcntc
\def\@citex[#1]#2{\if@filesw\immediate\write%
                  \@auxout{\string\citation{#2}}\fi
  \@tempcnta\z@\@tempcntb\m@ne\def\@citea{}\@cite{\@for\@citeb:=#2\do
    {\@ifundefined
       {b@\@citeb}{\@citeo\@tempcntb\m@ne\@citea%
                   \def\@citea{,}{\bf ?}\@warning
       {Citation `\@citeb' on page \thepage \space undefined}}%
    {\setbox\z@\hbox{\global\@tempcntc0\csname b@\@citeb%
                     \endcsname\relax}%
     \ifnum\@tempcntc=\z@ \@citeo\@tempcntb\m@ne
       \@citea\def\@citea{,}\hbox{\csname b@\@citeb\endcsname}%
     \else
      \advance\@tempcntb\@ne
      \ifnum\@tempcntb=\@tempcntc
      \else\advance\@tempcntb\m@ne\@citeo
      \@tempcnta\@tempcntc\@tempcntb\@tempcntc\fi\fi}}\@citeo}{#1}}
\def\@citeo{\ifnum\@tempcnta>\@tempcntb\else\@citea\def\@citea{,}%
  \ifnum\@tempcnta=\@tempcntb\the\@tempcnta\else
   {\advance\@tempcnta\@ne\ifnum\@tempcnta=\@tempcntb%
     \else \def\@citea{--}\fi
    \advance\@tempcnta\m@ne\the\@tempcnta\@citea\the\@tempcntb}\fi\fi}
\makeatother
\begin{document}
%
\vspace*{-1.0cm}
\begin{center}
{\Large\bf Pairing Properties In Relativistic Mean Field Models
Obtained From Effective Field Theory}
\\[2.0cm]
M. Del Estal, M. Centelles, X. Vi\~nas and S.K. Patra \\[2mm]
{\it Departament d'Estructura i Constituents de la Mat\`eria,
     Facultat de F\'{\i}sica,
\\
     Universitat de Barcelona,
     Diagonal {\sl 647}, E-{\sl 08028} Barcelona, Spain}
\end{center}
%
\vspace*{2.0cm}
\begin{abstract}
We apply recently developed effective field theory nuclear models in
mean field approximation (parameter sets G1 and G2) to describe
ground-state properties of nuclei from the valley of $\beta$-stability
up to the drip lines. For faster calculations of open-shell nuclei we
employ a modified BCS approach which takes into account quasi-bound
levels owing to their centrifugal barrier, with a constant pairing
strength. We test this simple prescription by comparing with available
Hartree-plus-Bogoliubov results. Using the new effective parameter
sets we then compute separation energies, density distributions and
spin--orbit potentials in isotopic (isotonic) chains of nuclei with
magic neutron (proton) numbers. The new forces describe the
experimental systematics similarly to conventional non-linear
$\sigma-\omega$ relativistic force parameters like NL3.
\end{abstract}

\mbox{}


\pagebreak

\section{Introduction}

The relativistic field theory of hadrons known as quantum
hadrodynamics (QHD) has become a very useful tool for describing bulk
and single-particle properties of nuclear matter and finite nuclei in
the mean field approximation \cite{Se86,Re89,Se92,Se97}. Compared with
the non-relativistic approach to the nuclear many-body problem, the
relativistic model explicitly includes the mesonic degrees of freedom
and treats the nucleons as Dirac particles. At the mean field
(Hartree) level, nucleons interact in a relativistic covariant way by
exchanging virtual mesons: an isoscalar-vector $\omega$ meson, an
isoscalar-scalar $\sigma$ meson and an isovector-vector $\rho$ meson.
With these ingredients the mean field treatment of QHD automatically
takes into account the spin--orbit force, the finite range and the
density dependence of the nuclear force. Adjusting some coupling
constants and meson masses from the properties of a small number of
finite nuclei, the relativistic mean field (RMF) model produces
excellent results for binding energies, root-mean-square radii,
quadrupole and hexadecapole deformations and other properties of
spherical and deformed nuclei \cite{Ga90,Ri97}.

The original linear $\sigma-\omega$ model of Walecka \cite{Wa74} was
complemented with cubic and quartic non-linearities of the $\sigma$
meson \cite{Bo77} (non-linear $\sigma-\omega$ model) to improve the
results for the incompressibility and for finite nuclei. Since these
models were proposed to be renormalizable, the scalar
self-interactions were limited to a quartic polynomial and
scalar-vector or vector-vector interactions were not allowed.
Recently, and inspired by effective field theory (EFT), Furnstahl,
Serot and Tang \cite{Fu96,Fu97} abandoned the idea of
renormalizability and extended the RMF theory by including other
non-linear scalar-vector and vector-vector self-interactions as well
as tensor couplings \cite{Se97,Fu96,Fu97,Mu96,Fu97b,Fu00}.

The EFT Lagrangian has an infinite number of terms since it contains
all the non-renormalizable couplings consistent with the underlying
QCD symmetries. Therefore, it is mandatory to develop a suitable
scheme of expansion and truncation. At normal nuclear densities the
scalar ($\Phi$) and vector ($W$) meson fields are small compared with
the nucleon mass ($M$), and they vary slowly with position in finite
nuclei. This indicates that the ratios $\Phi/M$, $W/M$,
$|\mbox{\boldmath $\nabla$}\Phi|/M^2$ and $|\mbox{\boldmath $\nabla$}
W|/M^2$ can be used as the expansion parameters. With the help of the
concept of naturalness, it is then possible to compute the
contributions of the different terms in the expansion and to truncate
the effective Lagrangian at a given level of accuracy
\cite{Se97,Fu97,Fu97b,Fu00}. None of the couplings should be
arbitrarily dropped out to the given order without a symmetry
argument. 

References \cite{Fu97,Fu97b,Fu00} have shown that it suffices to go to
fourth order in the expansion. At this level one recovers the standard
non-linear $\sigma-\omega$ model plus a few additional couplings, with
thirteen free parameters in all. These parameters have been fitted
(parameter sets G1 and G2) to reproduce some observables of magic
nuclei \cite{Fu97}. The fits display naturalness (i.e., all coupling
constants are of the order of unity when written in appropriate
dimensionless form), and the results are not dominated by the last
terms retained. This evidence confirms the utility of the EFT concepts
and justifies the truncation of the effective Lagrangian at the first
lower orders.

Recent applications of the models based on EFT include studies of
pion-nucleus scattering \cite{Cl98} and of the nuclear spin-orbit
force \cite{Fu98}, as well as calculations of asymmetric nuclear
matter at finite temperature with the G1 and G2 sets \cite{Wa00}. In a
previous work \cite{Es99} we have analyzed the impact of each one of
the new couplings introduced in the EFT models on the nuclear matter
saturation properties and on the nuclear surface properties. In Ref.\
\cite{Es00} we have looked for constraints on the new parameters by
demanding consistency with DBHF calculations and the properties of
finite nuclei. During the last years a large amount of work has been
devoted to measuring masses of nuclei far from stability \cite{Sa95}.
This body of experimental data has been used as a benchmark to test
the predictions of the currently existent (relativistic and
non-relativistic) nuclear effective forces \cite{Pa99}. This fact
motivates us to investigate in the present work the behaviour of the
parameter sets G1 and G2 derived from EFT in regions far from the
stability line.

To study ground-state properties of spherical open-shell nuclei one
has to take into account the pairing correlations. Relativistic mean
field calculations near the $\beta$-stability line have usually
included pairing in a constant gap BCS approximation
\cite{Ga90,Re86,Su94}, with the gaps fitted to empirical odd-even mass
differences. This approach works properly when the main effect of the
pairing correlations is a smearing of the Fermi surface. Since the BCS
pairing energy diverges for large momenta, a cut-off has to be
introduced in the pairing channel to simulate phenomenologically the
finite range of the particle-particle force. The limitations of this
simple BCS method appear when one deals with nuclei far from the
$\beta$-stability line. Close to the drip lines the Fermi level falls
near the particle continuum and it is known that the BCS model does
not provide a correct description of the coupling between bound and
continuum states \cite{Do84,Do96}. In the non-relativistic framework
this difficulty was overcome by the unified description of the mean
field and the pairing correlations provided by the
Hartree--Fock--Bogoliubov (HFB) theory \cite{Ei72,Ri80}, with Skyrme
\cite{Do84,Do96} or Gogny forces \cite{De80}.

The same unified treatment was developed by Kucharek and Ring
\cite{Ku91} in the relativistic framework. However, a quantitative
description of the pairing correlations in nuclei cannot be achieved
with relativistic mean field parametrizations because the meson
exchange forces are not properly adapted to large momentum transfer
\cite{Ku91,Go96}. Later, Ring and coworkers \cite{Go96,La98,Vr98,Sh00}
have used the RMF interaction for the particle-hole channel plus the
pairing part of the Gogny force \cite{De80} (with the D1S parameters
\cite{Be84}) for the particle-particle channel, in relativistic
Hartree-plus-Bogoliubov (RHB) calculations. Other authors have
employed a density-dependent zero-range pairing force \cite{Te95}
instead of the Gogny pairing force \cite{Me98,Me99}.

Recent calculations with non-relativistic Skyrme forces and a
zero-range force in the particle-particle channel have shown that a
BCS approach is able to provide a good qualitative estimate of the
drip lines if some quasibound states due to their centrifugal barrier
(plus the Coulomb barrier for protons) are included in the calculation
\cite{To79,Ch98,Sa00}. In this work we will use a similar BCS approach
with quasibound states to approximately take into account the effects
of the continuum contributions near the drip lines. We will employ a
constant pairing strength which can be considered as a simplification
of the zero-range pairing force and which gives similar results to
those obtained with a delta force for spherical nuclei \cite{Kr90}.

The paper is organized as follows. We summarize the mean field
approximation to the EFT nuclear model in the second section. In the
third section we describe our modified BCS approach with quasibound
states, and perform some calculations to test its possibilities and
limitations by comparing with Bogoliubov results available from the
literature. The fourth section is devoted to the detailed study with
the EFT parametrizations G1 and G2 of properties such as separation
energies, particle densities and spin--orbit potentials of nuclei
belonging to chains of isotopes (isotones) with magic proton (neutron)
number. Our conclusions are laid in the last section.

\section{Relativistic mean field approach from effective field theory}

The effective field theory approach to QHD has been developed in the
recent years. The theory and the equations for nuclear matter and
finite nuclei can be found in the literature \cite{Se97,Fu96,Fu97} and
here we shall only outline the formalism. We start from Ref.\
\cite{Fu96} where the field equations were derived from an energy
density functional containing Dirac baryons and classical scalar and
vector mesons. This functional can be obtained from the effective
Lagrangian in the Hartree approximation, but it can also be considered
as an expansion in terms of the ratios of the meson fields and their
gradients to the nucleon mass of a general energy density functional
that contains the contributions of correlations within the spirit of
density functional theory \cite{Se97,Fu97}.

According to Refs.\ \cite{Se97,Fu97} the energy density for finite
nuclei can be written as
\begin{eqnarray}
{\cal E}({\bf r}) & = &  \sum_\alpha \varphi_\alpha^\dagger
\Bigg\{ -i \mbox{\boldmath$\alpha$} \!\cdot\! \mbox{\boldmath$\nabla$}
+ \beta (M - \Phi) + W
+ \frac{1}{2}\tau_3 R
+ \frac{1+\tau_3}{2} A 
\nonumber \\[3mm]
& &
- \frac{i}{2M} \beta \mbox{\boldmath$\alpha$}\!\cdot\!
  \left( f_v \mbox{\boldmath$\nabla$} W
+ \frac{1}{2}f_\rho\tau_3 \mbox{\boldmath$\nabla$} R 
+ \lambda \mbox{\boldmath$\nabla$} A \right)
+ \frac{1}{2M^2}\left (\beta_s + \beta_v \tau_3 \right ) \Delta
A \Bigg\} \varphi_\alpha 
\nonumber \\[3mm]
& & \null
+ \left ( \frac{1}{2}
+ \frac{\kappa_3}{3!}\frac{\Phi}{M}
+ \frac{\kappa_4}{4!}\frac{\Phi^2}{M^2}\right )
 \frac{m_{s}^2}{g_{s}^2} \Phi^2  -
\frac{\zeta_0}{4!} \frac{1}{ g_{v}^2 } W^4 
\nonumber \\[3mm]
& & \null + \frac{1}{2g_{s}^2}\left( 1 +
\alpha_1\frac{\Phi}{M}\right) \left(
\mbox{\boldmath $\nabla$}\Phi\right)^2
 - \frac{1}{2g_{v}^2}\left( 1 +\alpha_2\frac{\Phi}{M}\right)
\left( \mbox{\boldmath $\nabla$} W  \right)^2
\nonumber \\[3mm]
& &  \null - \frac{1}{2}\left(1 + \eta_1 \frac{\Phi}{M} +
\frac{\eta_2}{2} \frac{\Phi^2 }{M^2} \right)
 \frac{{m_{v}}^2}{{g_{v}}^2} W^2  
- \frac{1}{2g_\rho^2} \left( \mbox{\boldmath $\nabla$} R\right)^2
- \frac{1}{2} \left( 1 + \eta_\rho \frac{\Phi}{M} \right)
\frac{m_\rho^2}{g_\rho^2} R^2
\nonumber \\[3mm]
& & \null
- \frac{1}{2e^2}\left( \mbox{\boldmath $\nabla$} A\right)^2
+ \frac{1}{3g_\gamma g_{v}}A \Delta W
+ \frac{1}{g_\gamma g_\rho}A \Delta R ,
\label{eqFN1}
\end{eqnarray}                                                          
where the index $\alpha$ runs over all occupied states $\varphi_\alpha
({\bf r})$ of the positive energy spectrum, $\Phi \equiv g_{s}
\phi_0({\bf r})$, $ W \equiv g_{v} V_0({\bf r})$, $R \equiv
g_{\rho}b_0({\bf r})$ and $A \equiv e A_0({\bf r})$. Variation of
the energy density (\ref{eqFN1}) with respect to
$\varphi^\dagger_\alpha$ and the meson fields gives the Dirac equation
fulfilled by the nucleons and the meson field equations, which are
solved self-consistently by numerical iteration. We refer the reader
to Ref.\ \cite{Fu97} for the expressions of the variational equations.

The terms with $g_\gamma$, $\lambda$, $\beta_{s}$ and $\beta_{v}$
take care of effects related with the electromagnetic structure of the
pion and the nucleon (see Ref.\ \cite{Fu97}). Specifically, the
constant $g_\gamma$ concerns the coupling of the photon to the pions
and the nucleons through the exchange of neutral vector mesons. The
experimental value is $g_\gamma^2/4\pi = 2.0$. The constant $\lambda$
is needed to reproduce the magnetic moments of the nucleons. It is
defined by
\begin{eqnarray}
\lambda =
\frac{1}{2} \lambda_{p} (1 + \tau_3) + \frac{1}{2} \lambda_{n} 
(1 - \tau_3) , 
\label{eqFN2} 
\end{eqnarray}
with $\lambda_{p} = 1.793$ and $\lambda_{n}=-1.913$ the anomalous
magnetic moments of the proton and the neutron, respectively. The
terms with $\beta_{s}$ and $\beta_{v}$ contribute to the charge
radii of the nucleon \cite{Fu97}.

In this work we will employ the EFT parameter sets G1 and G2 of Refs.\
\cite{Se97,Fu97}. The masses of the nucleon and the $\omega$ and
$\rho$ mesons take their experimental values: $M= 939$ MeV, $m_{v}=
782$ MeV and $m_\rho= 770$ MeV\@. The thirteen parameters $m_{s}$,
$g_{s}$, $g_{v}$, $g_\rho$, $\eta_1$, $\eta_2$, $\eta_\rho$,
$\kappa_3$, $\kappa_4$, $\zeta_0$, $f_{v}$, $\alpha_1$ and
$\alpha_2$ were fitted by a least-squares optimization procedure to
twenty-nine observables (binding energies, charge form factors and
spin--orbit splittings near the Fermi surface) of the nuclei $^{16}$O,
$^{40}$Ca, $^{48}$Ca, $^{88}$Sr and $^{208}$Pb, as described in Ref.\
\cite{Fu97}. The constants $\beta_{s}$, $\beta_{v}$ and $f_\rho$
were then chosen to reproduce the experimental charge radii of the
nucleon. The fits yielded two best, distinct parameter sets (G1 and
G2) with essentially the same $\chi^2$ value \cite{Fu97}.

We report in Table~1 the values of the parameters and the saturation
properties of G1 and G2. One observes that the fitted parameters
differ significantly between both interactions. For example, G2
presents a positive value of $\kappa_4$, as opposite to G1 and to many
of the most successful RMF parametrizations, such as the NL3 parameter
set \cite{La97}. Formally a negative value of $\kappa_4$ is not
acceptable because the energy spectrum then has no lower bound
\cite{Ba60}. Furthermore, the wrong sign in the $\Phi^4$ coupling
constant may cause troubles in obtaining stable solutions in light
nuclei like $^{12}$C\@. We note that the value of the effective mass
at saturation $M^*_\infty/M$ in the EFT sets ($\sim 0.65$) is somewhat
larger than the usual value in the RMF parameter sets ($\sim 0.60$).
This fact is related with the presence of the tensor coupling
$f_{v}$ of the $\omega$ meson to the nucleon, which has an important
bearing on the spin--orbit force \cite{Fu97,Fu98,Es99}.

One should mention that the EFT perspective also has been helpful to
elucidate the empirical success of the usual non-linear
$\sigma-\omega$ models that incorporate less couplings (just up to
cubic and quartic self-interactions of the scalar field): the EFT
approach accounts for the success of these RMF models and provides an
expansion scheme at the mean field level and for going beyond it
\cite{Se97,Fu97,Fu97b}. In practice it has been seen that the mean
field phenomenology of bulk and single-particle nuclear observables
does not constrain all of the new parameters of the EFT model
unambiguosly. That is, the constants of the EFT model are
underdetermined by the observables currently included in the fits and
different parameter sets with low $\chi^2$ (comparable to G1 and G2)
can be found \cite{Fu97,Fu97b,Fu00,Cl98}. However, the extra couplings
could prove to be very useful for the description of further
observables. Indeed, for densities above the normal saturation
density, and owing to the additional non-linear couplings, the EFT
models are able \cite{Es00} to give an equation of state and nuclear
matter scalar and vector self-energies in much better agreement with
the microscopic Dirac--Brueckner--Hartree--Fock predictions than the
standard non-linear $\sigma-\omega$ parametrizations (the latter
completely fail in following the DBHF trends as the nuclear density
grows \cite{Es00,Su94}).

The sets G1 and G2 were fitted including centre-of-mass corrections in
both the binding energy and the charge radius. Therefore, we will
utilize the same prescription of Ref.\ \cite{Fu97} in our
calculations with G1 and G2. Namely, a correction
\begin{eqnarray}
 E_{CM} = \frac{17.2}{A^{1/5}} \; {\rm MeV}
\label{eqFN22} 
\end{eqnarray}
to the binding energy and a correction
\begin{eqnarray}
 - \frac{3}{4} \frac{1}{(2 M A E_{CM})} \; {\rm fm^2}
\label{eqFN25} 
\end{eqnarray}
to the mean-square charge radius.

\section{The pairing calculation}
                     
It is well known that pairing correlations have to be included in any
realistic calculation of medium and heavy nuclei. In principle the
microscopic HFB theory should be used for this purpose. However, for
pairing calculations of a broad range of nuclei not too far from the
$\beta$-stability line, a simpler procedure is usually considered in
which a seniority potential acts between time-reversed orbitals. In
this section we want to discuss and test a straightforward improvement
of this simple approximation to be able to describe in addition nuclei
near the drip lines, at least on a qualitative level. Without the
complications intrinsic to a full Bogoliubov calculation, our faster
approximation will allow us later on to perform extensive calculations
of chains of isotopes and isotones with the relativistic parameter
sets. 

The pairing correlation will be considered in the BCS approach
\cite{Ei72,Ri80}. One assumes that the pairing interaction
$v_{\rm pair}$ has non-zero matrix elements only between pairs of
nucleons invariant under time reversal:
\begin{eqnarray}
\langle \alpha_2 \tilde\alpha_2|v_{\rm pair}|\alpha_1 \tilde\alpha_1
\rangle = -G \,,
\label{eqFN26}
\end{eqnarray}
where $|\alpha \rangle =| n l j m \rangle$ and $| \tilde\alpha \rangle
= | n l j \!-\! m \rangle$ (with $G>0$ and $m > 0$). Most often the
BCS calculations in the RMF model have been performed using a
constant gap approach \cite{Ga90,Re86,Su94}. Instead, here we choose a
seniority-type interaction with a constant value of $G$ for pairs
belonging to the active pairing shells.

The contribution of the pairing interaction to the total energy, for
each kind of nucleon (neutrons or protons), is
\begin{eqnarray}
E_{\rm pair} = -G \Bigg\{ \sum_{\alpha>0}
\left[n_{\alpha}(1-n_{\alpha})\right]^{1/2}
\Bigg\}^2-G\sum_{\alpha>0}n^2_{\alpha} \,,
\label{eqFN27}
\end{eqnarray}
where $n_\alpha$ is the occupation probability of a state with quantum
numbers $\alpha\equiv \{nljm\}$ and the sum is restricted to positive
values of $m$. One has
\begin{eqnarray}
n_\alpha = \frac{1}{2} \left[ 1 - \frac{\varepsilon_\alpha - \mu}
{\sqrt{
(\varepsilon_\alpha - \mu)^2 + \Delta^2}} \right] .
\label{eqFN28}
\end{eqnarray}
The Lagrange multiplier $\mu$ is called the chemical potential and the
gap $\Delta$ is defined by
\begin{eqnarray}
\Delta = 
G \sum_{\alpha} \left[n_{\alpha}(1-n_{\alpha})\right ]^{1/2} .
\label{eqFN29}
\end{eqnarray}
As usual the last term in Eq.\ (\ref{eqFN27}) will be neglected. It is
not a very important contribution and its only effect is a
renormalization of the pairing energies \cite{Ei72,Ri80}.

Assuming constant pairing matrix elements (\ref{eqFN26}) in the
vicinity of the Fermi level one gets \cite{Ei72,Ri80}
\begin{eqnarray}
\frac{G}{2} \sum_{\alpha > 0 } \frac{1} {\sqrt{
(\varepsilon_\alpha - \mu)^2 + \Delta^2}} & = & 1 ,
\label{eqFN30} \\[3mm]
 \frac{1}{2} \left[ 1- \frac{\varepsilon_\alpha - \mu}
 {\sqrt{(\varepsilon_\alpha - \mu)^2 + \Delta^2}} \right ] & = & A ,
\label{eqFN31}
\end{eqnarray}
where $A$ is the number of neutrons or protons involved in the pairing
correlation. The solution of these two coupled equations allows one to
find $\mu$ and $\Delta$. Using Eqs.\ (\ref{eqFN28}) and (\ref{eqFN29})
the pairing energy for each kind of nucleon can be written as
\begin{eqnarray}
E_{\rm pair} = -\frac{\Delta^2}{G} .
\label{eqFN32}
\end{eqnarray}

This simple approach breaks down for nuclei far from the stability
line. The reason is that in this case the number of neutrons (for
isotopes) or protons (for isotones) increases, the corresponding Fermi
level approaches zero and the number of available levels above it is
clearly reduced. Moreover, in this situation the particle-hole and
pair excitations reach the continuum. Ref.\ \cite{Do84} showed that if
one performs a BCS calculation using the same quasiparticle states as
in a HFB calculation, then the BCS binding energies are close to the
HFB ones but the r.m.s.\ radii (i.e., the single-particle wave
functions) dramatically depend on the size of the box where the
calculation is performed. This is due to the fact that there are
neutrons (protons) that occupy continuum states for which the wave
functions are not localized in a region, thus giving rise to an
unphysical neutron (proton) gas surrounding the nucleus.

Recent non-relativistic calculations near the drip lines with Skyrme
forces \cite{Ch98,Sa00} have shown that the above problem of the BCS
approach can be corrected, in an approximate manner, by taking into
account continuum effects by means of the so-called quasibound states,
namely, states bound because of their own centrifugal barrier
(centrifugal-plus-Coulomb barrier for protons). When the quasibound
states are included in the BCS calculation (from now on a qb-BCS
calculation), it is necessary to prevent the unrealistic pairing of
highly excited states and to confine the region of influence of the
pairing potential to the vicinity of the Fermi level. Instead of using
a cutoff factor as in Ref.\ \cite{Ch98}, in our calculations we will
restrict the available space to one harmonic oscillator shell above
and below the Fermi level.

In order to check this approach we have performed with the G1
parameter set ($G_{n}= 21/A$ MeV, see next section) calculations of
the binding energy and r.m.s.\ radius of the $^{120}$Sn and $^{160}$Sn
nuclei in boxes of sizes between 15 and 25 fm (as in the
non-relativistic calculations of Ref.\ \cite{Do84}). The results
taking into account the quasibound levels $1h_{9/2}$, $2f_{5/2}$ and
$1i_{13/2}$ for $^{120}$Sn, and $1i_{11/2}$ and $1j_{13/2}$ for
$^{160}$Sn, are compared in Figure~1 with the output of a standard BCS
calculation with only bound levels. It turns out that in the qb-BCS
case the results are essentially independent of the size of the box
where the calculations are carried out. When the quasibound levels are
included the binding energies are larger than when only the bound
levels are taken into account, due to the damping of the pairing
correlation caused by disregarding the continuum states in the
standard BCS calculation \cite{Do84}. We also show in Figure~1 the
results of a BCS calculation using all bound and unbound levels (i.e.,
without restricting ourselves to quasibound levels) in the considered
range. It is obvious that in this case the results are box dependent,
as the binding energy and neutron r.m.s.\ radius of $^{160}$Sn
evidence. 

Another test of the qb-BCS approach concerns the asymptotic behaviour
of the particle densities \cite{Do96}. In Figure~2 we display the
radial dependence of the neutron density of $^{150}$Sn (as in Ref.\
\cite{Do96}) calculated with the G1 parameter set in boxes of radii
between 15 and 25 fm. For large enough distances the density decreases
smoothly when the size of the box increases (except very near of the
edge, where the density suddenly drops to zero because of the
$\varphi_\alpha=0$ boundary condition). This means that no neutron gas
surrounding the nucleus has appeared. In a Bogoliubov calculation the
asymptotic behaviour of the particle density is governed by the square
of the lower component of the single-quasiparticle wave function
corresponding to the lowest quasiparticle energy \cite{Do96}. This
asymptotic behaviour is displayed by the (almost straight) dotted line
in Figure~2. It can be seen that the density obtained with our
approach decreases more slowly than the RHB density, i.e.,
asymptotically the qb-BCS density is not able to follow the RHB
behaviour. This coincides with the conclusion of Ref.\ \cite{Do96}
(see Figure~19 of that work) where non-relativistic HFB densities are
compared for large distances with the densities obtained in the qb-BCS
approach with a state-dependent pairing \cite{To79}.

Although the qb-BCS densities do not display the right asymptotic
behaviour, it was conjectured in Ref.\ \cite{Do96} that such an
approach could allow one to compute properties of nuclei much closer
to the drip lines than in a standard BCS calculation. Very recently,
RHB calculations up to the drip lines of the two-neutron separation
energy $S_{2n}$ for nickel isotopes \cite{Me98} and of the charge
and neutron r.m.s.\ radii for tin isotopes \cite{Me99} have been
carried out using the NL-SH parameter set \cite{Sh93} plus a
density-dependent zero-range pairing force. We have repeated these
calculations with our qb-BCS method for both isotopic chains (with a
pairing interaction strength $G_{n}= 22.5/A$ in the case of NL-SH).

We display the values of the $S_{2n}$ separation energies for the Ni
chain in Figure~3a. The RHB calculation predicts the drip line at the
isotope $^{100}$Ni and shows shell effects at $N=28$ and 50 (and to a
minor extent at $N=70$). These features are well predicted by our
simpler qb-BCS calculation. The differences between these qb-BCS and
RHB results also come in part from the different pairing forces used
in the calculations. To investigate this point we show in Figure~3b
the neutron pairing energy obtained in our approach [Eq.\
(\ref{eqFN32})] for the isotopes of the Ni chain. It vanishes at $N=
28$, 50 and 70, in agreement with the shell structure shown in
Figure~3a by the $S_{2n}$ separation energies. The largest pairing
energies are found in the middle of two closed shells and they are
enhanced by increasing $N$. Figure~3b can be compared with the RHB
values displayed in Figure~2 of Ref.\ \cite{Me98}. The tendencies are
the same, though the qb-BCS pairing energies are slightly larger than
in the RHB calculation. In Figure~4 we draw our results for the radii
of the Sn isotopes, and compare them with the RHB values. In the case
of the charge radii the agreement is excellent. The neutron radii
obtained in our method closely follow the behaviour of the RHB neutron
radii and the kink at $N=132$ is qualitatively reproduced.

We have furthermore computed the binding energies of nuclei of the
$N=20$ isotonic chain for which RHB results exist with the NL3
parameter set \cite{Vr98}. We present the extracted two-proton
separation energies $S_{2p}$ in Table~2. The agreement between the
qb-BCS and RHB approaches again is very good. In both models the last
stable nucleus is $^{46}$Fe, as in experiment. Notice that in the
present case the first levels with positive energy correspond to those
of the $pf$ shell. Due to the Coulomb barrier all these levels become
quasibound in our approach, and it is expected that they will lie
close to the canonical levels. This explains the goodness of the
qb-BCS energies for this isotonic chain.

>From the previous comparisons we see that the simple qb-BCS
calculation is able to reasonably follow the main trends of the more
fundamental RHB pairing calculation. One can also conclude that the
consideration of quasibound states in the BCS approach is, actually, a
key ingredient to eliminate the spurious nucleon gas arising near the
drip lines.

\section{Results for EFT parameter sets}

We want to analyze the ability of the G1 and G2 parameter sets based
on effective field theory \cite{Se97,Fu97} to describe nuclear
properties far from the stability line, i.e., far from the region
where the parameters were fitted. To our knowledge such calculations
have not been explored so far. We will contrast the results with
experiment and with those predicted by the NL3 set, that we take as
one of the best representatives of the usual RMF model with only
scalar self-interactions.

As indicated, we shall use a schematic pairing with a
state-independent matrix element $G_{\tau} = C_{\tau}/A$, where
$C_\tau$ is a constant and $\tau= n, p$ for neutrons or
protons, respectively. We fix the constant $C_{n}$ for neutrons by
looking for the best agreement of our calculation with the known
experimental binding energies of Ni and Sn isotopes. Similarly, we
determine $C_{p}$ for protons from the experimental binding energies
of the isotones of $N=28$ and $N=82$. The values obtained from this
fit are $C_{n}= 21$ MeV and $C_{p}=22.5$ MeV for the G1 set,
$C_{n}=19$ MeV and $C_{p}=25$ MeV for G2, and finally $C_{n}=20.5$ MeV
and $C_{p}=23$ MeV for NL3. Figure~5 shows that the neutron and proton
state-independent gaps ($\Delta_{n}$ and $\Delta_{p}$) predicted by
our calculation with G1 are scattered around the empirical average
curve $12/\sqrt{A}$ \cite{Bo69}. A similar picture is found with the
parameter sets G2 and NL3.

\subsection{Two-particle separation energies}

In Figure~6a we present the two-neutron separation energies $S_{2n}$
for the chain of Ni isotopes. Clear shell effects arise at $N=28$ and
50. The three relativistic interactions (G1, G2 and NL3) slightly
overestimate the shell effect at $N=28$ as compared with the
experimental value, which also happens in more sophisticated RHB
calculations with NL3 \cite{La98,Sh00}. In our qb-BCS approach some
disagreement with experiment is found for the $N=38$ and $N=40$
isotopes. Again, this also occurs in the RHB calculations of Refs.\
\cite{La98,Sh00} with NL3. However, if we compare Figure~6a with the
results that we have shown in Figure~3a for the NL-SH parameter set,
we see that NL-SH achieves a better agreement with experiment for
these $N=38$ and $N=40$ isotopes.

We stop our calculation towards the neutron drip line when the
two-neutron separation energy vanishes or when the neutron chemical
potential becomes positive. The fact that $S_{2n}$ is not always
zero at the drip line is connected with the quenching of the shell
structure with $N$, which is a force-dependent property \cite{Do96}.
This effect is illustrated in Figure~25 of Ref.\ \cite{Do96} for HFB
calculations with different non-relativistic forces. We find similar
situations with the considered relativistic sets in our qb-BCS
calculations of separation energies. In the case of the Ni isotopes we
reach the drip line at $N=66$ with the G1 and NL3 sets and at $N=68$
with the G2 set. This agrees nicely with the value $N=66$ obtained in
HFB calculations with the Skyrme forces SIII \cite{Te95,Te96} and SkP
\cite{Te95}. For NL-SH our qb-BCS scheme predicts the drip line at
$N=72$ (see Figure~3a), the same value found in the RHB calculations
of Ref.\ \cite{Me98}.

In Figure~7a we display our qb-BCS results for the two-neutron
separation energies of the Sn isotopic chain. In Ref.\ \cite{Sh00} it
was claimed that pure BCS calculations in the constant gap approach
(with NL3) are not suitable for the Sn isotopes. We observe in
Figure~7a that below $N=60$, as one moves towards $N=50$, some
discrepancies with the experimental values appear, which also arise in
the RHB calculations \cite{Sh00}. The three forces slightly
overestimate the shell effect at $N=82$ (as the RHB results of Refs.\
\cite{La98,Sh00} for NL3). We have computed Sn isotopes up to $A=176$,
when $S_{2n}$ vanishes for NL3 (in good agreement with RHB results for
NL-SH \cite{Me99} and HFB results for the Skyrme force SkP
\cite{Do84}). For G1 and G2 we find that $S_{2n}$ does not yet vanish
at $N=126$, and it is not possible to increase the neutron number due
to the shell closure at $N=126$ (the neutron chemical potential
becomes positive for the $N=128$ isotope). This means that the
quenching of the shell effect at $N=126$ for NL3 (and NL-SH) is larger
than for the G1 and G2 parameter sets.

Our calculated $S_{2n}$ energies for Pb isotopes are shown in
Figure~8. The experimental shell effect at $N=126$ is reasonably well
reproduced by G1, G2 and NL3. The drip line is found at $N=184$ with a
non-vanishing two-neutron separation energy, as in the calculations
performed with the extended Skyrme force SLy4 in Ref.\ \cite{Ch98},
where a similar approach (quasibound states and a state-dependent gap)
to ours was used. The relatively large shell effect found at $N=184$
means that there is no quenching for this magic number in our qb-BCS
approximation for the studied parameter sets. Indeed, to verify this
point a full RHB calculation should be performed.

To analyze the proton pairing we have studied the two-proton
separation energies in chains of isotones of $N=28$ (Figure~9a) and
$N=82$ (Figure~10a). In the case of $N=28$, shell gaps appear at
$Z=20$ and $Z=28$. For $Z=20$ the predicted gap is larger than in
experiment. For $Z=28$, G1 and G2 agree better than NL3 with
experiment. $S_{2p}$ vanishes at $Z=30$ for NL3, whereas it vanishes
at $Z=32$ for G1 and G2. The isotones of $N=82$ display a clear shell
effect at $Z=50$, in agreement with the non-relativistic calculation
of Ref.\ \cite{Ch98}. It is slightly larger for G2 than for G1 and
NL3. Experimental information for this shell effect is not available.
NL3 would predict another shell effect at $Z=58$, which does not
appear experimentally. The effect is less pronounced in G1 and it does
not show up in G2. The three forces indicate that the proton drip line
is reached after the $^{156}$W isotope, in agreement with experimental
information \cite{Page92}.

Figures~11a and 11b show, respectively, the calculated $S_{2p}$
separation energies for the $N=50$ and $N=126$ isotone chains. Note
that we did not use any information about these nuclei in our fit of
the $G_{p}$ pairing strength. For $N=50$ the set G2 follows the
experimental data very well, specially for the larger $Z$. The trend
of G1 and NL3 is only a little worse. The proton drip line is located
at $^{100}$Sn in the three parametrizations, in good accordance with
experiment. The quenching of the shell effect at $Z=50$ is larger for
G2 than for G1 and NL3. The available data for two-proton separation
energies of $N=126$ isotones are reasonably well estimated by the
relativistic sets. However, the trend of NL3 is worse than that of G1
and G2. It would then be very interesting to perform RHB calculations
of this chain to confirm the behaviour of NL3. The last nucleus of the
chain stable against two-proton emission is $^{218}$U according to G1
and NL3, and $^{220}$Pu according to G2. The three sets predict a
shell effect at $Z= 92$, though it is relatively quenched for G2.

\subsection{One-particle separation energies}

We have computed one-neutron (one-proton) separation energies for Ni
and Sn isotopes (for $N=28$ and $N=82$ isotones). The results are
displayed in Figures~6b and 7b (9b and 10b), respectively. To deal
with odd mass number nuclei we have used a spherical blocking
approximation. One pair of conjugate states $|\alpha\rangle$ and
$|\tilde\alpha\rangle$ is blocked, i.e., taken out of the pairing
scheme \cite{Ei72,Ri80}. In the spherical approximation one replaces
the blocked single-particle state by an average over the degenerate
states in its $j$ shell. This way the rotational and time-reversal
invariance of the many-body system is restored in the intrinsic frame
\cite{Be00}. In this approach the contribution of the $j$ shell that
contains the blocked state to the number of active particles and the
pairing energy is
\begin{eqnarray}
A_j & = & (2j-1)n_j + 1 ,
\label{eqFN33} \\[3mm]
E_{{\rm pair} , j} & = & -G \, (2j-1) [n_j(1-n_j) ]^{1/2} ,
\label{eqFN34} 
\end{eqnarray}
respectively. The remaining active shells contribute in the usual
manner [Eqs.\ (\ref{eqFN27}) and (\ref{eqFN31})]. Due to rearrangement
effects, blocking the single-particle state with smallest
quasiparticle energy $E_{\alpha}={\sqrt{(\varepsilon_\alpha - \mu)^2 +
\Delta^2}} $ in the even $A-1$ nucleus, does not necessarily lead to
the largest binding energy of the odd $A$ nucleus. Therefore, in some
cases one has to repeat the calculation blocking in turn the different
single-particle states that lie around the Fermi level to find the
configuration of largest binding energy \cite{Do84,De80,Be00}.

The one-neutron (one-proton) separation energies lie over two
different curves for even and odd neutron (proton) number. For Ni
isotopes (Figure~6b) the three parameter sets G1, G2 and NL3 reproduce
reasonably the experimental values. The shell effect at $N=28$ is,
again, overestimated by the three forces. The heaviest Ni isotope
stable against one-neutron emission is found at $N=55$ with NL3 and G1
and at $N=57$ with G2. For Sn isotopes (Figure 7b) the shell effect at
$N=82$ is slightly overestimated by the studied forces. The
predictions of the three parametrizations are roughly similar up to
$N=110$, where the behaviour of NL3 starts to depart from G1 and G2
due to the large quenching of the $N=126$ shell effect shown by NL3 as
compared with G1 and G2. In our calculations, the odd Sn isotopes
become unstable against the emission of one neutron around $N=110$.
This value is larger than the values found with the non-relativistic
SkP interaction in HFB ($N=103$) or HF+BCS ($N=101$) calculations
\cite{Do84}. The origin of this discrepancy lies in the fact that the
shell distribution in tin isotopes around $N=126$ for SkP is rather
different from that of the relativistic sets \cite{Sh00}.

The one-proton separation energies for the isotones of $N=28$
(Figure~9b) show an overall good agreement with the experimental data.
The shell effects at $Z=20$ and $Z=28$ are rather well reproduced by
the forces analyzed here. G1 and G2 predict the heaviest nucleus
stable against one-proton emission to be $^{57}$Cu, as in experiment,
while it is unstable in the NL3 calculation. For the isotones of
$N=82$ (Figure~10b) the shell effect predicted by NL3 and G1 at $Z=50$
is similar. Again, as for $S_{2p}$ (Figure~10a), NL3 predicts a
shell effect at $Z=58$ which is not found experimentally, whereas for
G1 this effect is clearly smaller and it does not appear for G2. The
last stable nucleus against one-proton emission is $^{151}$Tm
according to the three parameter sets.

\subsection{One-body densities and potentials}

The nuclear densities for chains of isotopes of light and medium size
nuclei have recently been studied in the RHB approximation
\cite{La98,Vr98,Me98,Me99}. As $N$ grows the neutron and mass
densities extend outwards and the r.m.s.\ radii and the surface
thickness increase. Special attention has been payed to the isospin
dependence of the spin--orbit interaction. The magnitude of the
spin--orbit potential is reduced when one approaches the neutron drip
line and, as a consequence, there is a reduction of the energy
splittings between spin--orbit partner levels \cite{La98,Vr98,Me99}.
To our knowledge, for isotones such an study has only been carried out
in the $N=20$ chain \cite{Vr98}. It is to be remarked that the EFT
parametrizations G1 and G2 contain a tensor coupling of the $\omega$
meson to the nucleon which plays a very important role in the
spin--orbit force because there exists a trade-off between the size of
this coupling and the size of the scalar field \cite{Fu98,Es99}.

In Figures~12a and 12b we display, respectively, the neutron and
proton densities of some $N=28$ isotones from $Z=16$ to $Z=32$ as
predicted by the G2 set in our qb-BCS approach. Figures~13a and 13b
show the results for some $N=82$ isotones from $Z=40$ to $Z=70$. Since
$N$ is fixed in each isotonic chain, the spatial extension of the
neutron densities is very similar for the different nuclei of the
chain. In any case, as one goes from the lightest to the heaviest
isotone of the chain, the neutron densities tend to be depressed in
the interior region and their surface thickness (90\%--10\% fall-off
distance) shows a decreasing tendency. The proton densities of the
isotones exhibit a strong dependence on $Z$: by adding more protons
they are raised at the interior and their surface is pushed outwards.
For $N=28$ the surface thickness of $\rho_{p}$ remains roughly
constant up to $Z=28$ and increases for heavier isotones as a
consequence of the growing occupation of the $1f_{7/2}$ shell. At the
origin the proton densities show a bump when $Z\ge 20$ because the
$2s_{1/2}$ level is occupied. The $Z=16$ isotone shows a dip at the
centre due to, precisely, the emptiness of this $2s_{1/2}$ level. For
the considered nuclei of $N=82$, the proton densities have an
approximately constant surface thickness and present a hole at the
centre owing to the Coulomb repulsion. In Figure~14 we display the
neutron and proton r.m.s.\ radii of the $N=82$ isotonic chain obtained
with G2 and NL3. It turns out that the predictions of both sets are
very similar. The proton radii increase uniformly with $Z$, similarly
to the behaviour found for $N=20$ isotones with NL3 and the RHB scheme
in Ref.\ \cite{Vr98}. The neutron radii remain roughly constant with
$Z$. They just show a slight decrease with increasing $Z$ till $Z\sim
50$ and slightly increase afterwards. This behaviour may be related
with the shell effect for protons at $Z=50$.

The spin--orbit interaction is automatically included in the RMF
approximation. It appears explicitly when the lower spinor of the
relativistic wave function is eliminated in favour of the upper
spinor. This way one obtains a Schr\"odinger-like equation with a term
$V_{SO} (r)$ that has the structure of the single-particle
spin--orbit potential. Including the contribution of the tensor
coupling of the $\omega$ meson, the spin--orbit term reads
\cite{Fu98,La98}
\begin{eqnarray}
H_{SO} & = & 
\frac{1}{2 M^2} V_{SO}(r) \, {\bf L}\!\cdot\!{\bf S} ,
\label{eqSO2} \\[3mm]
V_{SO}(r) & = & \frac{M^2}{{\overline{M}}^2} \frac{1}{r}
\left( \frac{d \Phi}{d r} + \frac{d W}{d r} \right)
+ 2 f_{v} \frac{M}{\overline{M}} \frac{1}{r} \frac{d W}{d r} ,
\label{eqSO1}
\end{eqnarray}
where $\overline{M} = M - {\textstyle{1\over2}} (\Phi + W)$.
We have checked numerically that the contribution to the spin--orbit
potential of the $f_\rho$ tensor coupling of the $\rho$ meson is very
small, even when one approaches the drip lines. Hence we have not
written this contribution in Eq.\ (\ref{eqSO1}).

The spin--orbit potential (\ref{eqSO1}) for some lead isotopes
computed with G2 and NL3 is displayed in Figures~15a and 15b,
respectively. As a general trend, for both G2 and NL3, when the number
of neutrons is increased the depth of the spin--orbit potential
decreases gradually and the position of the bottom of the well is
shifted outwards, which implies a significant weakening of the
spin--orbit interaction. The same effect arises in other isotopic
chains in RHB calculations \cite{La98,Me98,Me99}. Comparing the
spin--orbit potentials obtained with the G2 and NL3 sets, one sees
that they have a similar strength for all the isotopes analyzed and
that the minima of the wells are located at similar positions
(slightly shifted to larger values of $r$ in G2). The higher effective
mass of G2 at saturation ($M^*_\infty/M=0.664$) with respect to NL3
($M^*_\infty/M=0.595$) is compensated by the tensor coupling included
in G2 ($f_{v}=0.692$). To ascertain the relative importance of the
tensor coupling we have drawn in the insert of Figure~15a, for
$^{228}$Pb, the full potential (\ref{eqSO1}) and the contribution
resulting from setting $f_{v}=0$ in Eq.\ (\ref{eqSO1}). We see that
the full $V_{SO}(r)$ is much deeper and wider. The maximum depth of
$V_{SO}(r)$ changes from $-68$ MeV~fm$^{-2}$ (right scale of the
insert) to $-44$ MeV~fm$^{-2}$ when $f_{v}=0$. That is, the tensor
coupling accounts for roughly one third of the total spin--orbit
strength in the G2 parameter set.

One expects that the weakening of the spin--orbit potential in going
to the neutron drip line will bring about a reduction of the
spin--orbit splittings
\begin{eqnarray}
 \Delta\varepsilon = \varepsilon_{n l, j=l-1/2} 
 - \varepsilon_{n l, j=l+1/2} 
\label{SO3}
\end{eqnarray}
of the neutron levels \cite{La98}. Figure~16 displays the energy
splittings of some spin--orbit partner levels of neutrons for lead
isotopes, obtained with the G2 and NL3 parameter sets. The splittings
predicted by G2 and NL3 are very close as a consequence of the
similarity of the corresponding spin--orbit potentials. Partner levels
with high angular momentum undergo some reduction in the splitting
along the Pb isotopic chain, but partners with small angular momentum
show an almost constant splitting. By comparison of their RHB results
for Ni and Sn, the authors of Ref.\ \cite{La98} pointed out that the
weakening of the spin--orbit interaction should be less important for
heavier isotopic chains. Our calculations for Pb would confirm this
statement. All the single-particle levels involved in Figure~16 are
bound. Of course, one should not expect the results for
$\Delta\varepsilon$ to be so reliable in our qb-BCS approach if one,
or both, of the partner levels lies at positive energy. The reason is
that the single-particle energies of the quasibound levels do not
exactly reproduce the energies of the corresponding canonical states
of a RHB calculation.

In Figures~17a and 17b we show the spin--orbit potential for isotones
of $N=82$ from $Z=40$ to $Z=70$, for the G2 and NL3 parametrizations.
Similarly to what is found for isotopes, the results obtained from G2
and from NL3 are comparable and the spin--orbit potential well
$V_{SO}(r)$ moves outwards with the addition of protons, following
the tendency of the proton density. However, for isotones we find that
the behaviour of the depth of the spin--orbit potential well is not so
monotonous: it increases when one goes from the neutron drip line up
to the $\beta$-stable region, while it decreases afterwards as more
protons are added.

\section{Summary and conclusion}

We have analyzed the pairing properties of some chains of isotopes and
isotones with magic $Z$ and $N$ numbers in the relativistic mean field
approach. The study has been performed for the G1 and G2
parametrizations that were obtained in Ref.\ \cite{Fu97} from the
modern effective field theory approach to relativistic nuclear
phenomenology. We have compared the results with those obtained with
the NL3 parameter set which is considered to be very successful for
dealing with nuclei beyond the stability line.

For accurate calculations of pairing far from the valley of
$\beta$-stability in the relativistic models, the relativistic
Hartree--Bogoliubov approach should be applied. However, we have
presented a simpler modified BCS approach which allows one to obtain
pairing properties near the drip lines fast and confidently. The
method has been used previously in non-relativistic calculations with
Skyrme forces \cite{Ch98,Sa00}. The key ingredient is to take into
account the continuum contributions through quasibound levels due to
their centrifugal barrier. To further simplify the calculations we
have assumed pairing matrix elements of the type $G=C/A$ instead of,
e.g., a state-dependent pairing with a zero-range force.

The considered quasibound levels are mainly localized in the
classically allowed region and decrease exponentially outside it. This
eliminates the unphysical nucleon gas which, near the drip lines,
surrounds the nucleus when all available positive energy levels are
included in the usual BCS approach. Normally, the quasibound levels
have high angular momentum and lie close in energy to the
corresponding RHB canonical levels. One of the limitations of the
qb-BCS approach employed here is the fact that the nuclear density
does not follow the asymptotic fall-off of the densities computed with
the relativistic Hartree-plus-Bogoliubov theory. In spite of this
shortcoming, we have shown by comparison with available RHB results
that the qb-BCS approach is able to predict the position of the drip
lines, or the behaviour of the neutron and charge radii for nuclei far
from the stability line, in a reasonable way. Also, the obtained
pairing gaps are nicely scattered around the empirical average
$12/\sqrt{A}$. 

We have applied the qb-BCS approach to the Ni, Sn and Pb isotopic
chains, and to the $N=28$, 50, 82 and 126 isotonic chains. The
two-neutron (two-proton) and one-neutron (one-proton) separation
energies, as well as the resulting shell gaps, are similar for the
three studied relativistic parametrizations (G1, G2 and NL3) and in
general they reproduce the available experimental data, at least
qualitatively. The neutron and proton drip lines are usually reached
at the same place with the three forces, though one may find a shift
of one or two units of $A$ among them. We have paid some attention to
the quenching of the shell structure near the drip lines. For example,
the quenching of the shell effect at $N=126$ for Sn isotopes is larger
in NL3 than in G1 and G2, while for Pb isotopes none of the three sets
exhibits a quenching of the shell effect at $N=184$ in our qb-BCS
calculation. The EFT parametrizations G1 and G2 contain tensor
couplings that are not present in the RMF parametrizations like NL3
and have a larger effective mass at saturation. However, the predicted
spin--orbit potentials along the isotopic and isotonic chains do not
differ much from those obtained with NL3.

Our analysis shows that the parameter sets based on EFT are able to
describe nuclei far from the $\beta$-stability line, after adding a
phenomenological pairing residual interaction. Only experimental
information about some magic nuclei was utilized in the fit of the
constants of the G1 and G2 sets and, thus, the results for nuclei near
the drip lines are veritable predictions of the model. In spite of the
fact that the EFT sets include more couplings and parameters than the
conventional RMF sets like NL3, both models reproduce the experimental
systematics with a similar quality. In fact, the studied properties
away from the valley of $\beta$-stability do not seem to provide
further constraints on the EFT parameters, not even in the isovector
sector. In conclusion, extended sets like G1 and G2 will serve almost
the same purposes for normal systems as the conventional parameter
sets. However, some of the extra parameters of the general EFT
functional may be used to better describe regions of the equation of
state at higher density or temperature \cite{Wa00,Es00} without
spoiling the systematics for finite nuclei.

\section{Acknowledgments}
The authors would like to acknowledge support from the DGICYT (Spain)
under grant PB98-1247 and from DGR (Catalonia) under grant
1998SGR-00011.  S.K.P. thanks the Spanish Education Ministry grant
SB97-OL174874 for financial support and the Departament d'Estructura i
Constituents de la Mat\`eria of the University of Barcelona for kind
hospitality.

\pagebreak


%
\pagebreak

%
%
\section*{Figure captions}
\begin{description}
\item[Figure 1.]
Dependence of the binding energy (left) and neutron r.m.s.\ radius
(right) of the nuclei $^{120}$Sn and $^{160}$Sn on the size of the box
used in the calculations. Solid, dashed and dotted lines correspond to
a BCS approach including quasi-bound levels, only bound levels and all
available levels, respectively. The results are for the G1 parameter
set. 
\item[Figure 2.]
Neutron density of $^{150}$Sn for different sizes of the box used in
the qb-BCS calculations (for the set G1). The dotted line denotes the
asymptotic behaviour expected from a Bogoliubov calculation
\cite{Do96}. 
\item[Figure 3.]
Part (a): the two-neutron separation energies for Ni isotopes
calculated in the qb-BCS approach are compared with the RHB results
of Ref.\ \cite{Me98} and with experiment. Part (b): the neutron
pairing energy obtained in the qb-BCS approach. The results are for
the set NL-SH.
\item[Figure 4.]
Charge and neutron r.m.s.\ radii of Sn isotopes in qb-BCS and RHB
\cite{Me99} calculations performed with the NL-SH set.
\item[Figure 5.]
The state-independent pairing gaps predicted by our qb-BCS approach
for Ni, Sn and Pb isotopes (top) and for $N=28$, 50, 82 and 126
isotones (bottom). The G1 set has been used. The empirical average
curve $12/\sqrt{A}$ \cite{Bo69} is depicted by a solid line.
\item[Figure 6.]
Two-neutron (a) and one-neutron (b) separation energies for Ni
isotopes computed with the qb-BCS approach for the parametrizations
G1, G2 and NL3, in comparison with the experimental data.
\item[Figure 7.]
Same as Figure~6 for Sn isotopes.
\item[Figure 8.]
Two-neutron separation energies for Pb isotopes computed with the
qb-BCS approach, in comparison with experiment.
\item[Figure 9.]
Two-proton (a) and one-proton (b) separation energies for $N=28$
isotones computed with the qb-BCS approach, in comparison with
experiment. 
\item[Figure 10.]
Same as Figure~9 for $N=82$ isotones.
\item[Figure 11.]
Two-proton separation energies for $N=50$ (a) and $N=126$ (b)
isotones computed with the qb-BCS approach, in comparison with
experiment. 
\item[Figure 12.]
Radial dependence of the neutron (a) and proton (b) densities of some
$N=28$ isotones obtained with the G2 set.
\item[Figure 13.]
Same as Figure~12 for some $N=82$ isotones.
\item[Figure 14.]
Neutron and proton r.m.s.\ radii of $N=82$ isotones obtained with
the G2 and NL3 sets.
\item[Figure 15.]
Spin--orbit potential for some Pb isotopes obtained with the G2 set
(a) and with the NL3 set (b).
\item[Figure 16.]
Energy splitting of some spin--orbit partner levels of neutrons in Pb
isotopes, calculated in the qb-BCS approach for the G2 and NL3 sets.
\item[Figure 17.]
Spin--orbit potential for some $N=82$ isotones obtained with the G2
set (a) and with the NL3 set (b).
\end{description}

\pagebreak

%
\section*{Tables}
\begin{table}
\caption{Dimensionless parameters and saturation properties of the
sets G1 and G2 based on EFT and of the RMF set NL3.}
\vspace*{1.cm}
\centering
\begin{tabular}{ccccccccc}
\hline
\hline
 & &  G1 &&  G2 && NL3 \\
\hline
$m_{s}/M$     & & 0.540 & & 0.554 & &  0.541\\
$g_{s}/4\pi$  & & 0.785 & & 0.835 & &  0.813\\
$g_{v}/4\pi$  & & 0.965 & & 1.016 & &  1.024\\
$g_\rho/4\pi$ & & 0.698 & & 0.755 & &  0.712\\
$\kappa_3$    & & 2.207 & & 3.247 & &  1.465\\
$\kappa_4$    & & $-$10.090 & & 0.632 & & $-$5.668\\
$\zeta_0$     & & 3.525 & & 2.642 & &    0.0\\
$\eta_1$      & & 0.071 & & 0.650 & &    0.0\\
$\eta_2$      & & $-$0.962 & & 0.110 & &    0.0\\
$\eta_\rho$   & & $-$0.272 & & 0.390 & &    0.0\\
$\alpha_1$    & & 1.855 & & 1.723 & &    0.0\\
$\alpha_2$    & & 1.788 & & $-$1.580 & &    0.0\\
$f_{v}/4$     & & 0.108 & & 0.173 & &    0.0\\
$f_\rho/4$    & & 1.039 & & 0.962 & &    0.0\\
$\beta_s$     & & 0.028 & & $-$0.093 & &    0.0\\
$\beta_v$     & & $-$0.250 & & $-$0.460 & &    0.0\\
\hline
$a_{v}$ (MeV) & & $-$16.14 & & $-$16.07 & &  $-$16.24\\
$\rho_\infty$ (fm$^{-3}$) & & 0.153 & & 0.153 & &  0.148\\
$K$ (MeV)     & & 215.0 & & 215.0 & & 271.5\\
$M^{*}_{\infty}/M$ & & 0.634 & & 0.664 & &  0.595\\
$J$ (MeV)     & &  38.5 & & 36.4 & &  37.40\\
\hline
\hline
\end{tabular}
\end{table}

\vspace*{1cm}

\begin{table}
\caption{RHB and qb-BCS two-proton separation energies (in MeV) of
some $N=20$ isotones calculated with the NL3 parameter set.}
\vspace*{1.cm}
\centering
\begin{tabular}{ccccccccc}
\hline
\hline
$S_{2p}$    & & RHB && qb-BCS && exp \\
\hline
$^{36}$S      & &  23.56 & & 23.05 & & 25.28\\
$^{38}$Ar     & &  19.36 & & 18.97 & & 18.35\\
$^{40}$Ca     & &  14.65 & & 15.46 & & 14.99\\
$^{42}$Ti     & &  6.36 & & 6.70 & & 4.86\\
$^{44}$Cr     & &  3.30 & & 3.31 & & 3.08\\
$^{46}$Cr     & &  0.60 & & 0.54 & & 0.21\\
$^{48}$Ni     & &  $-$2.33 & & $-$2.21 & & $-$\\
\hline
\hline
\end{tabular}
\end{table}

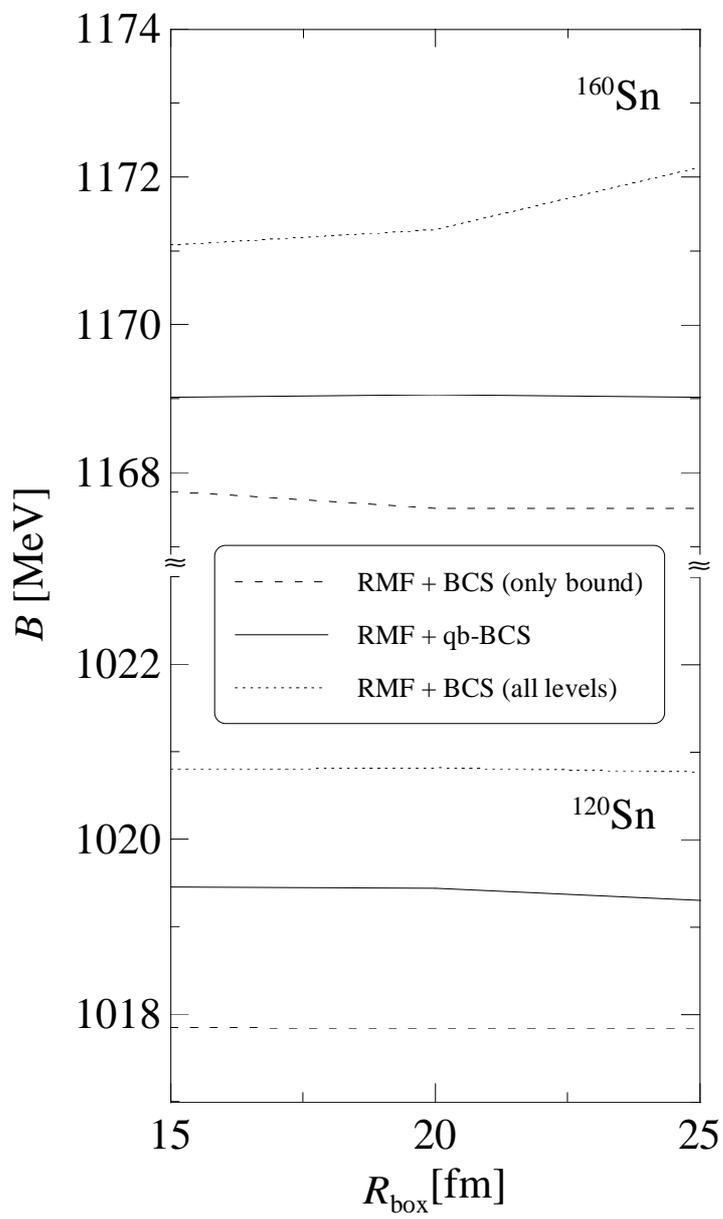
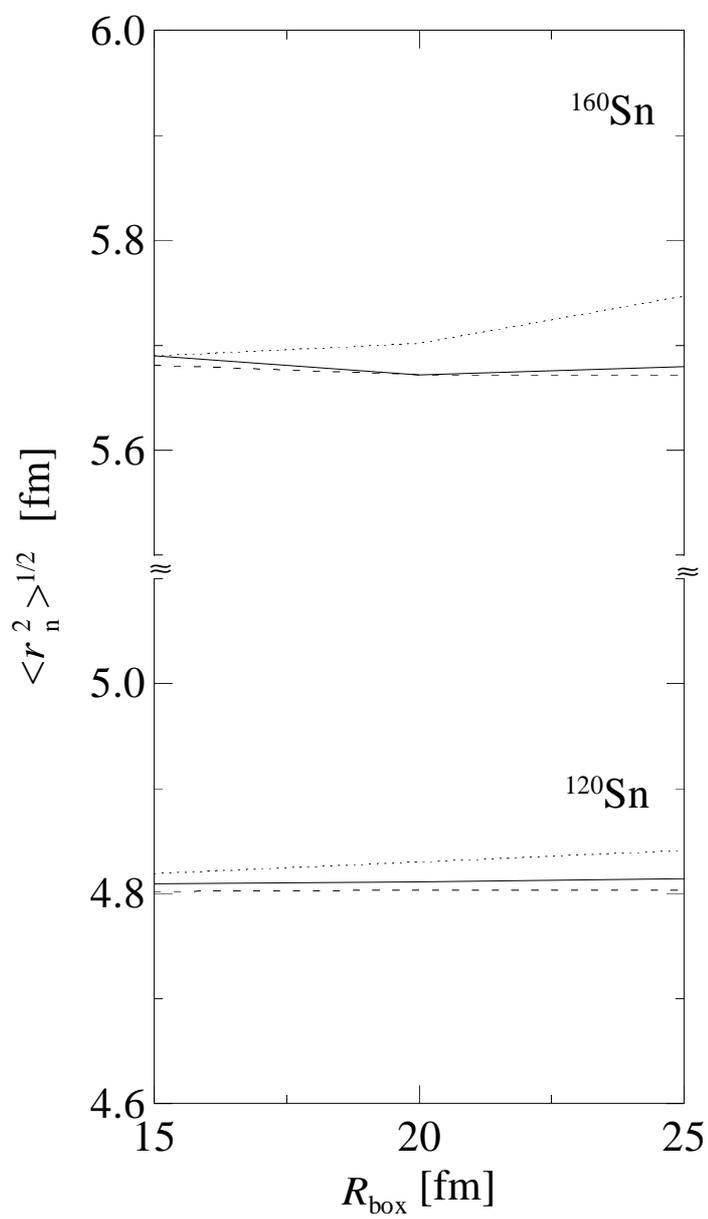
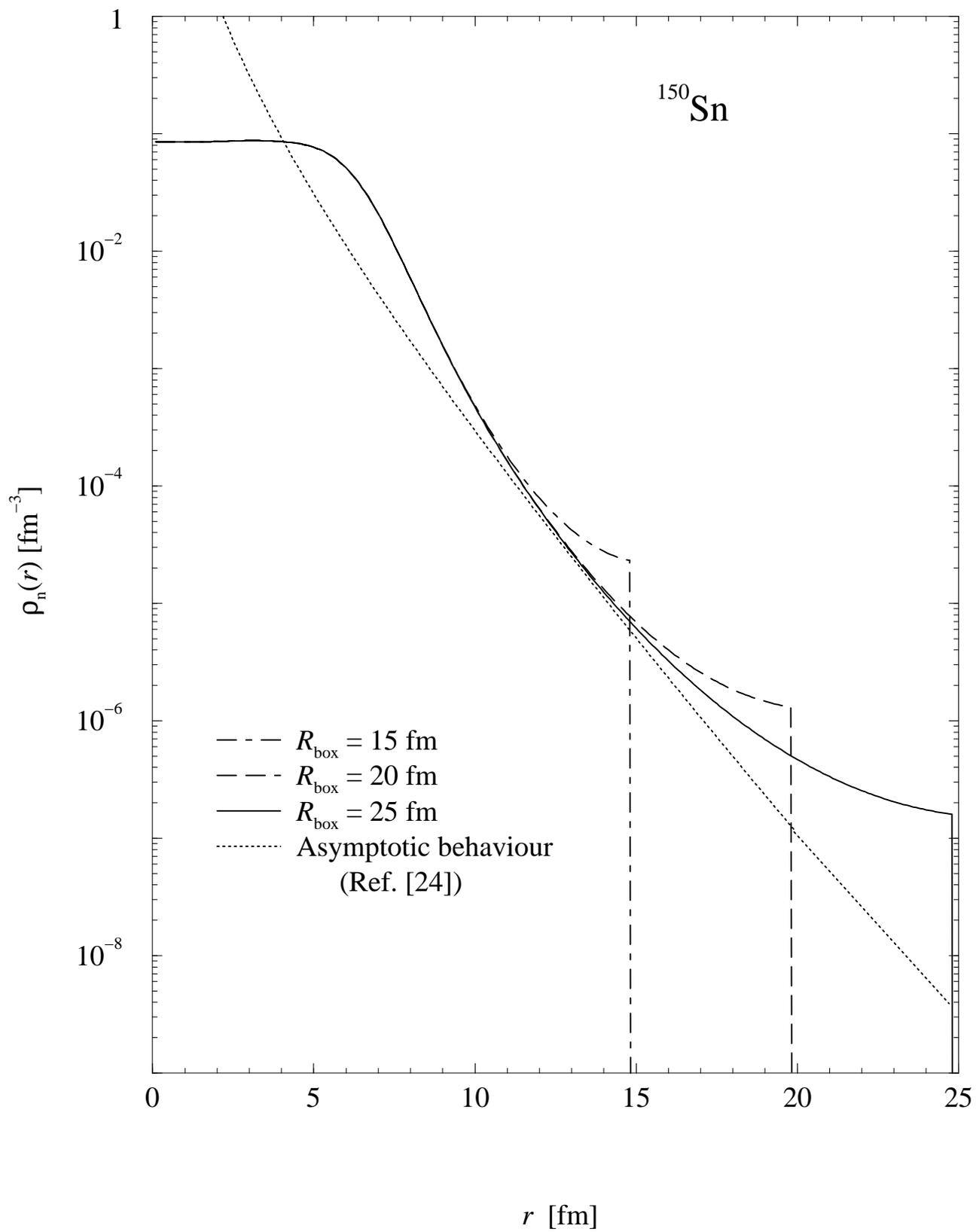
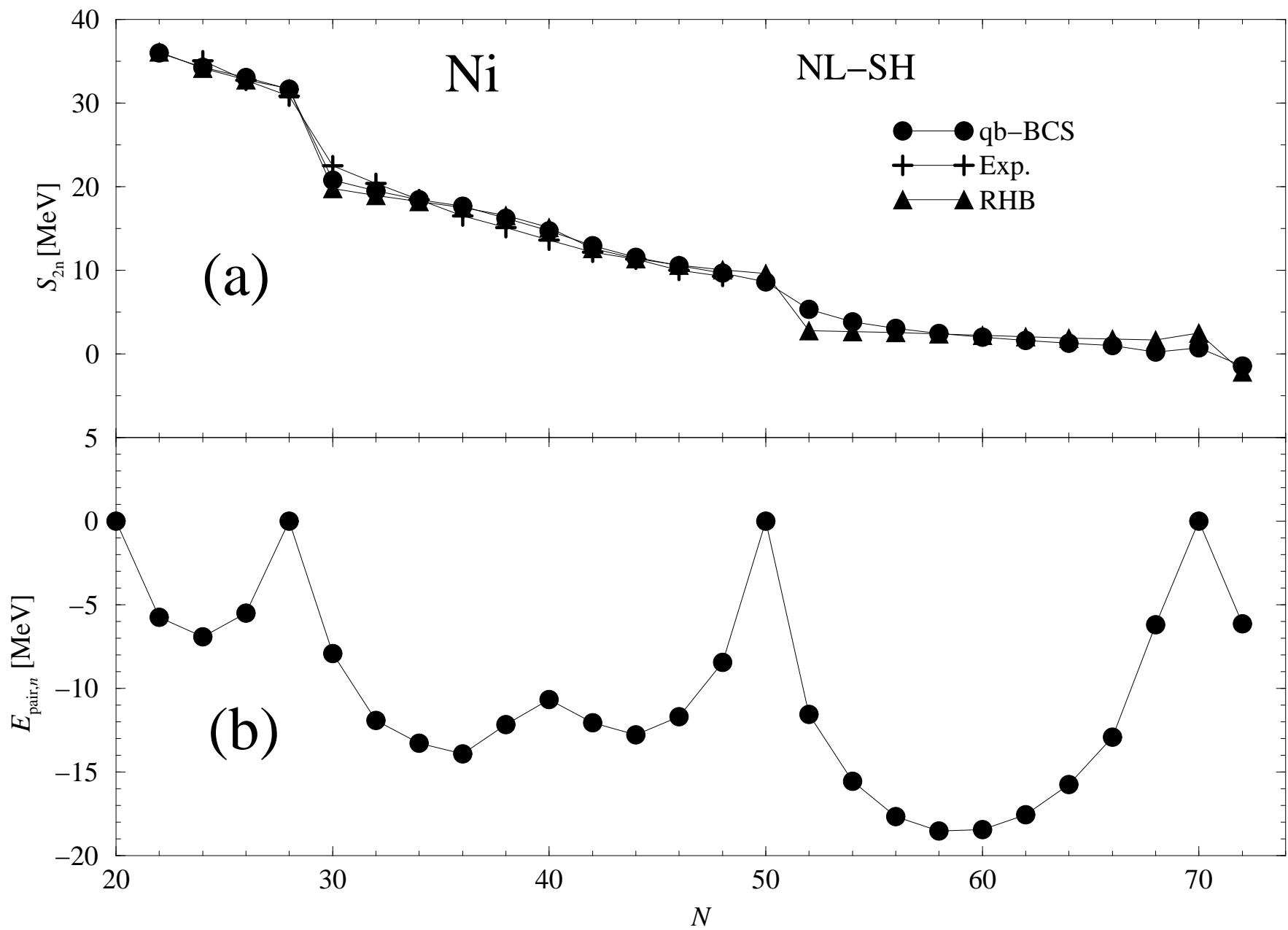
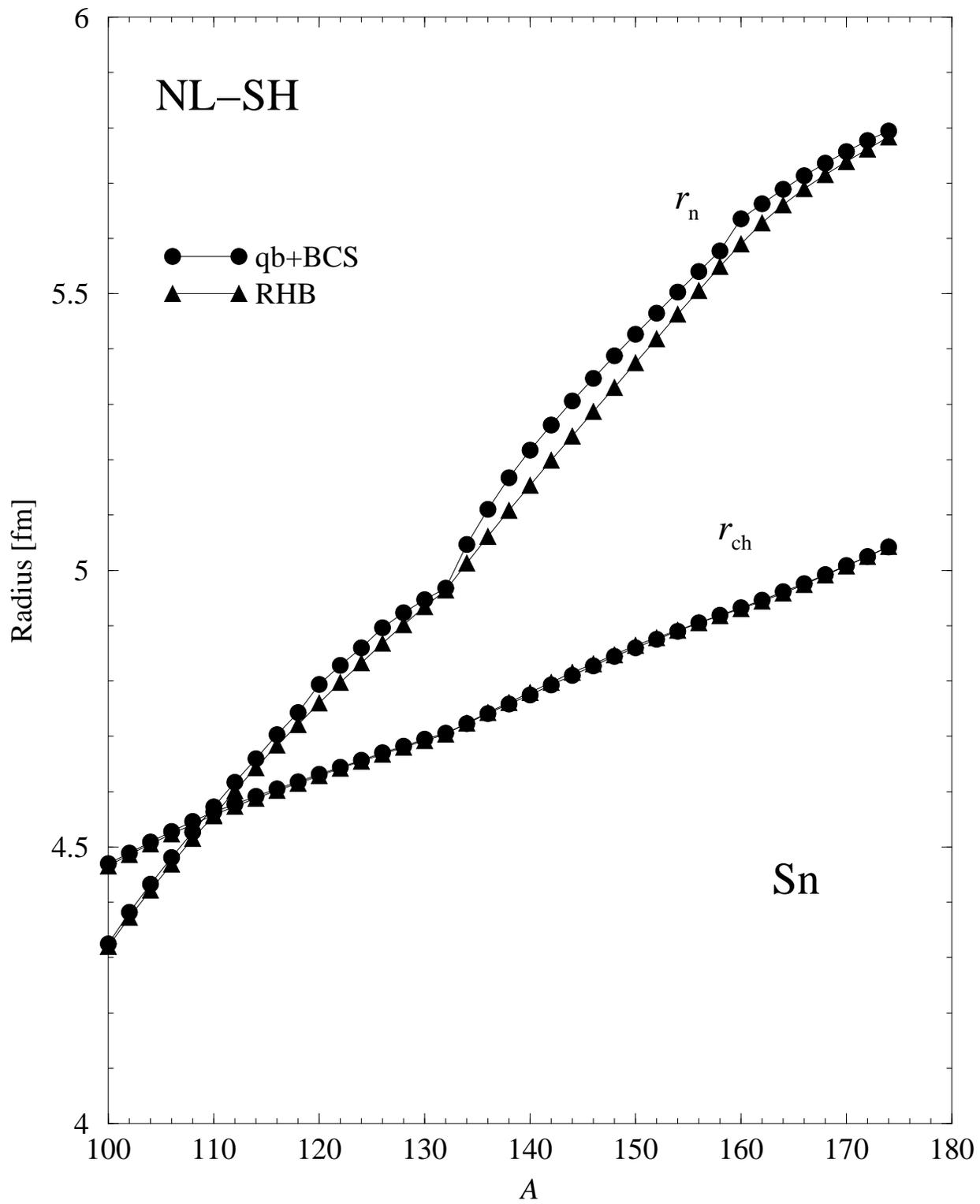
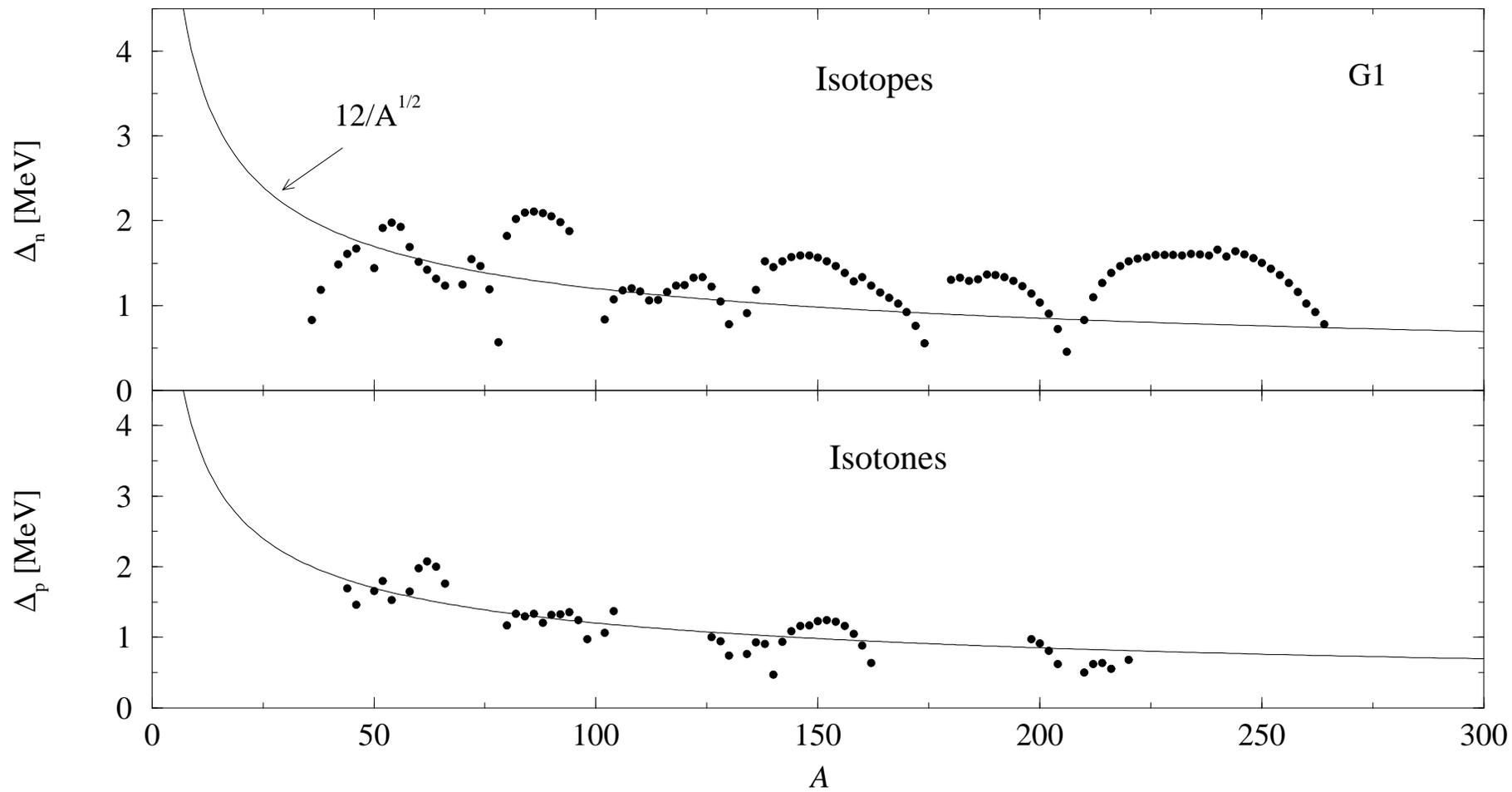
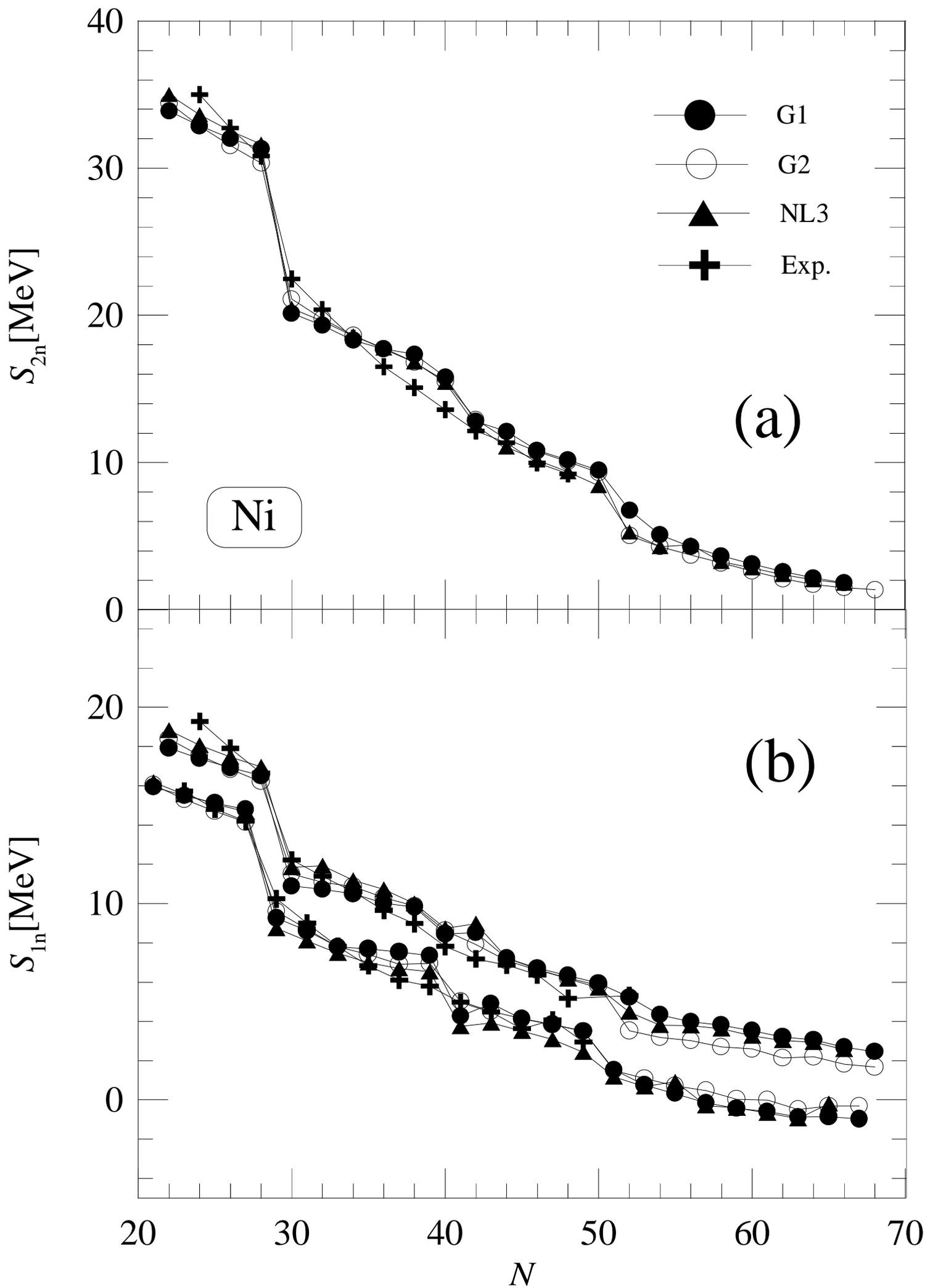
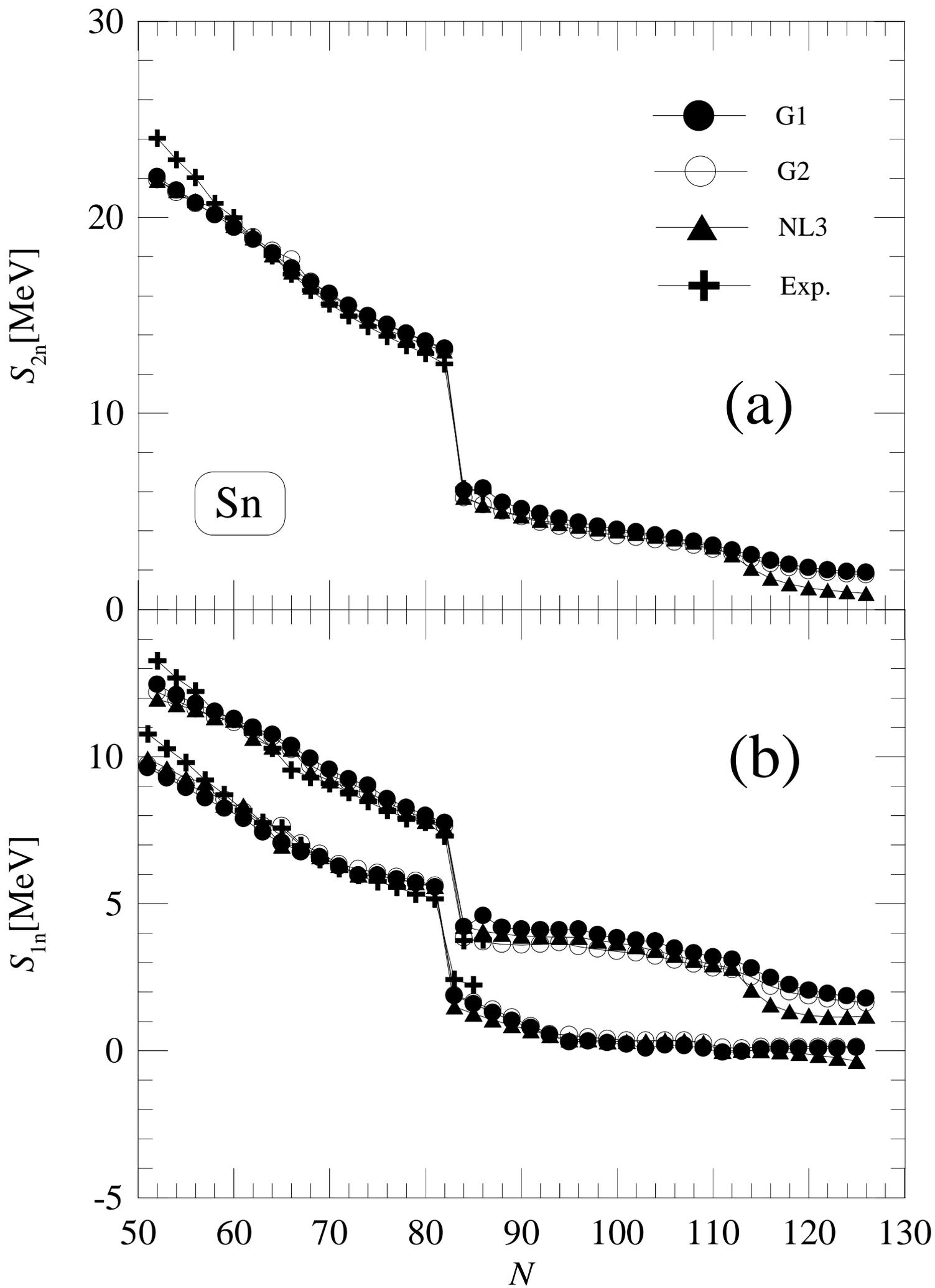
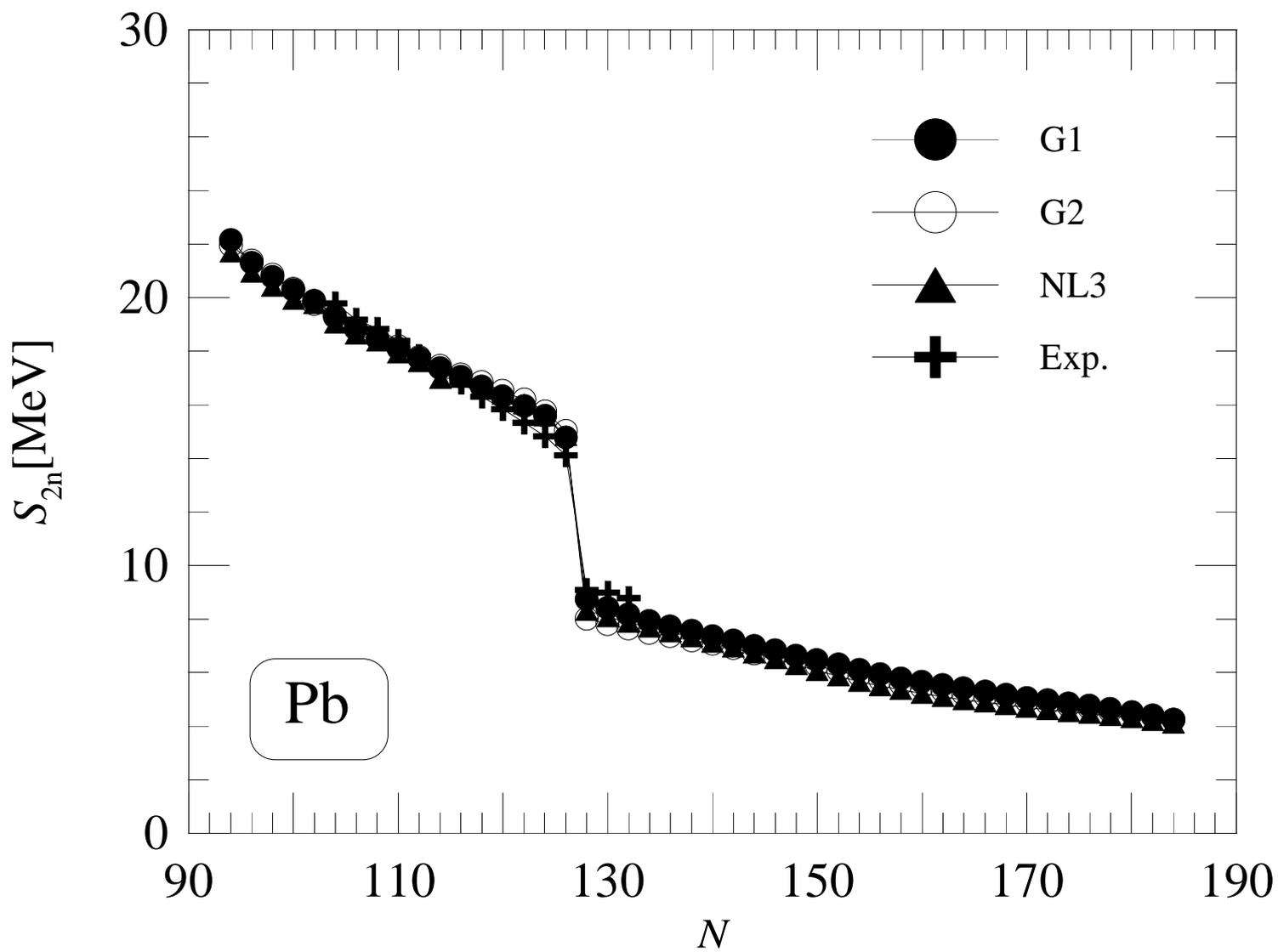
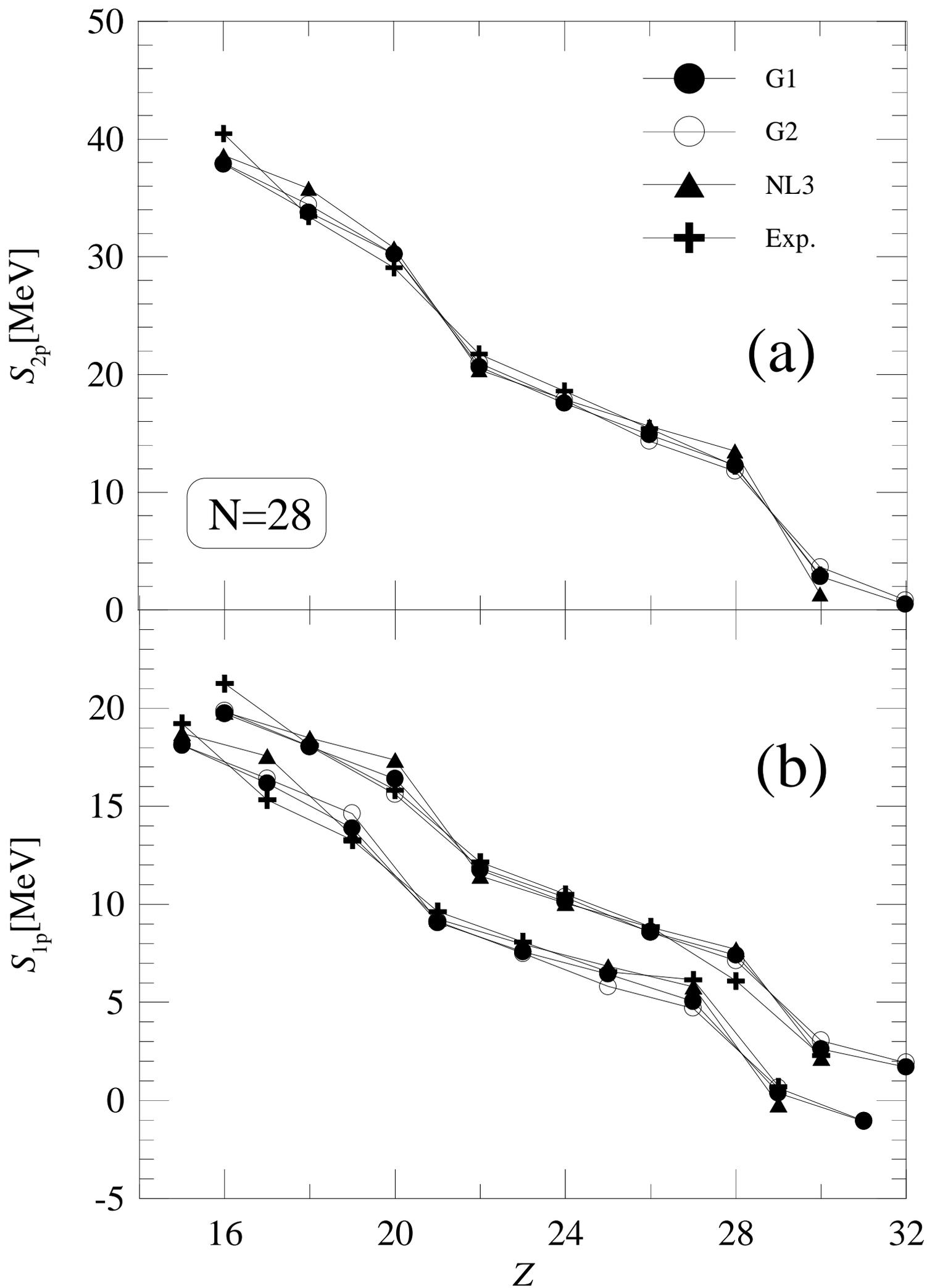
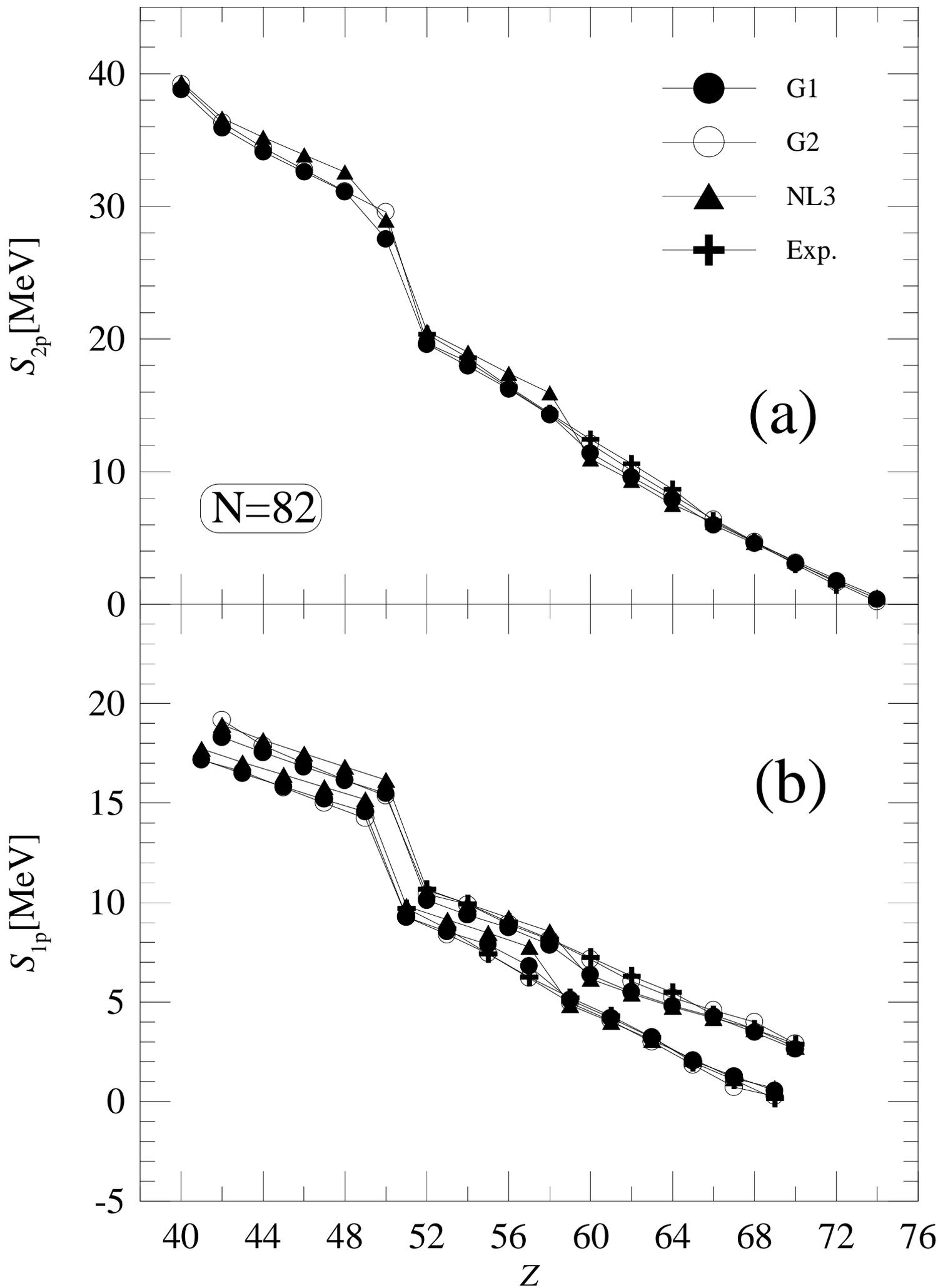
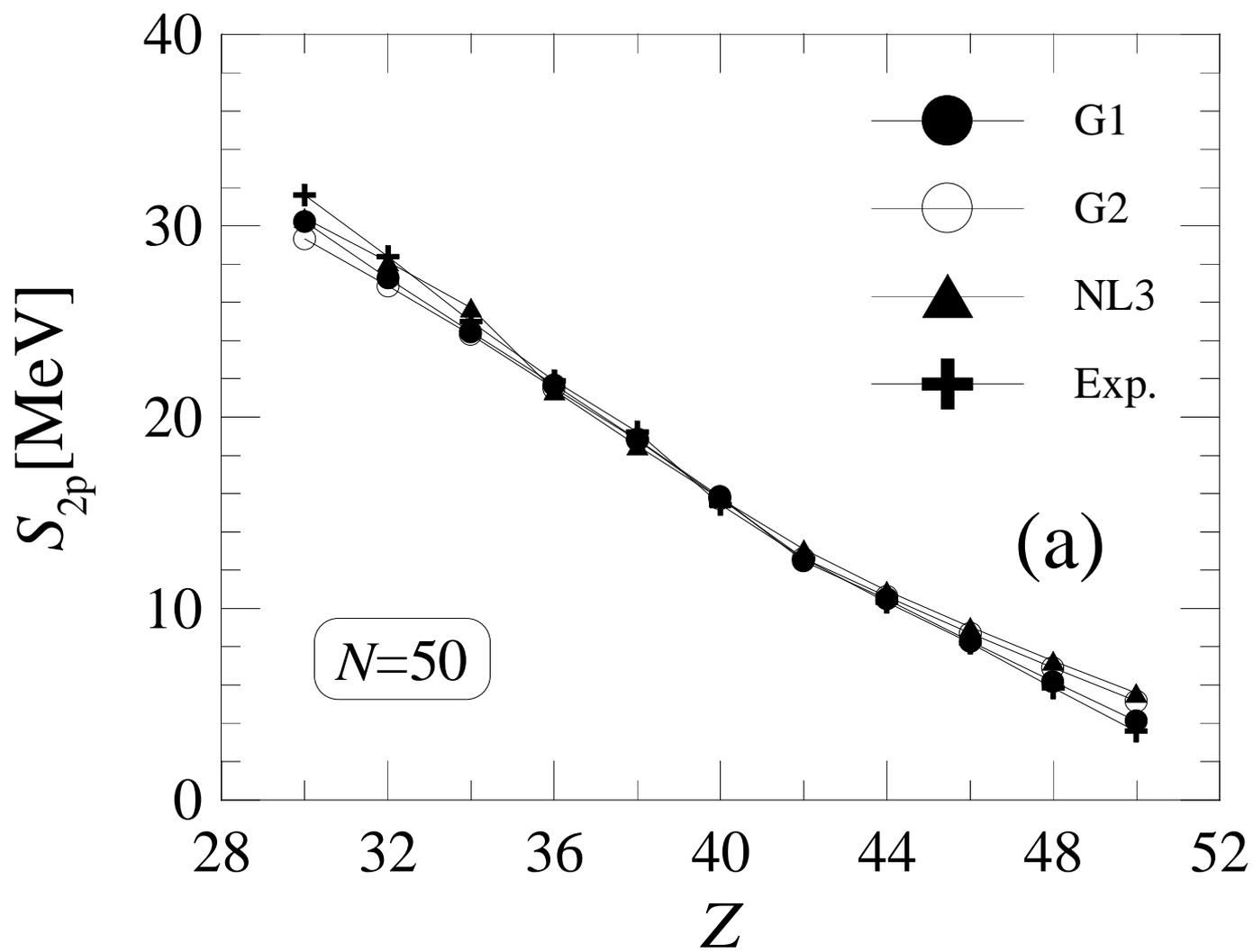
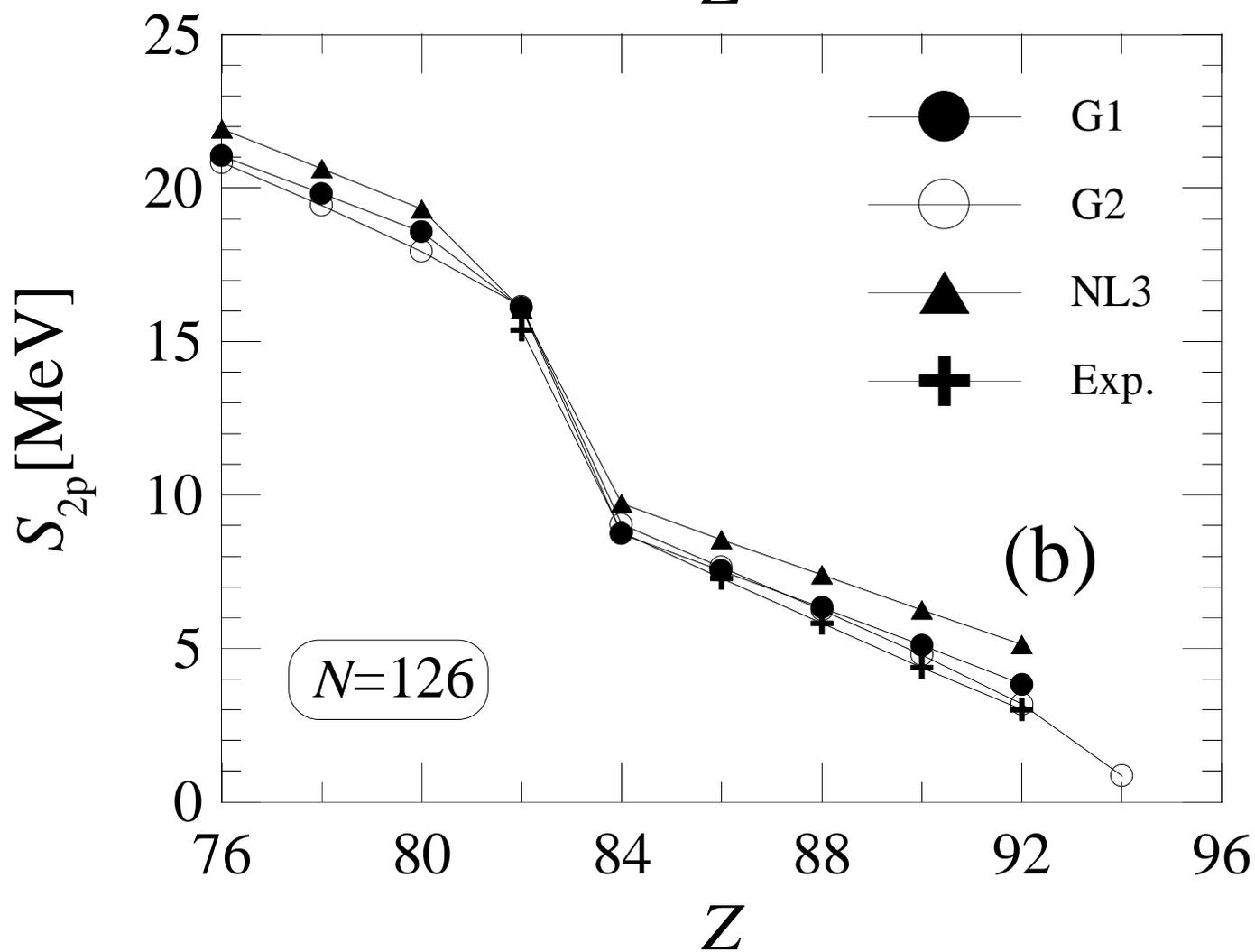
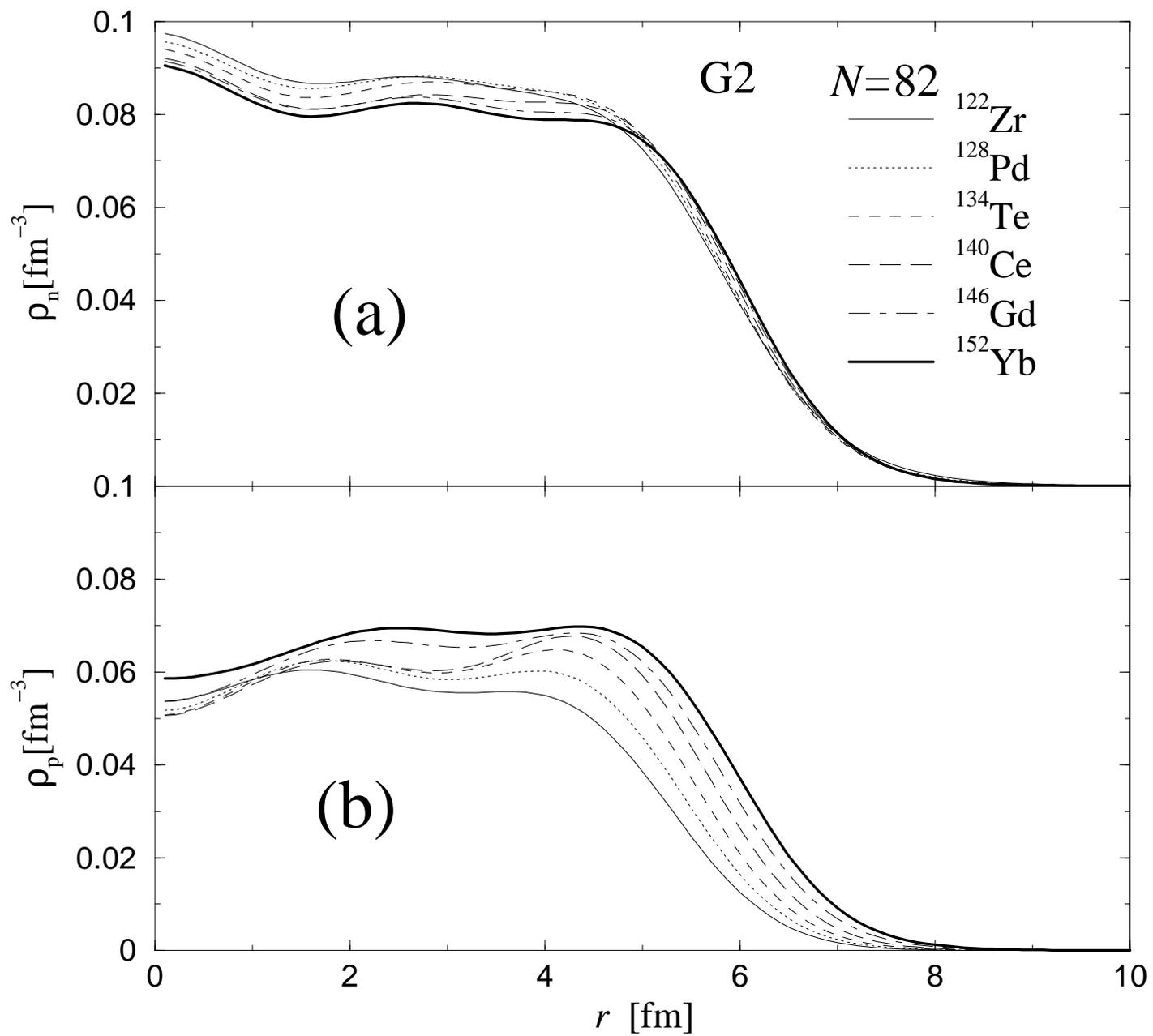
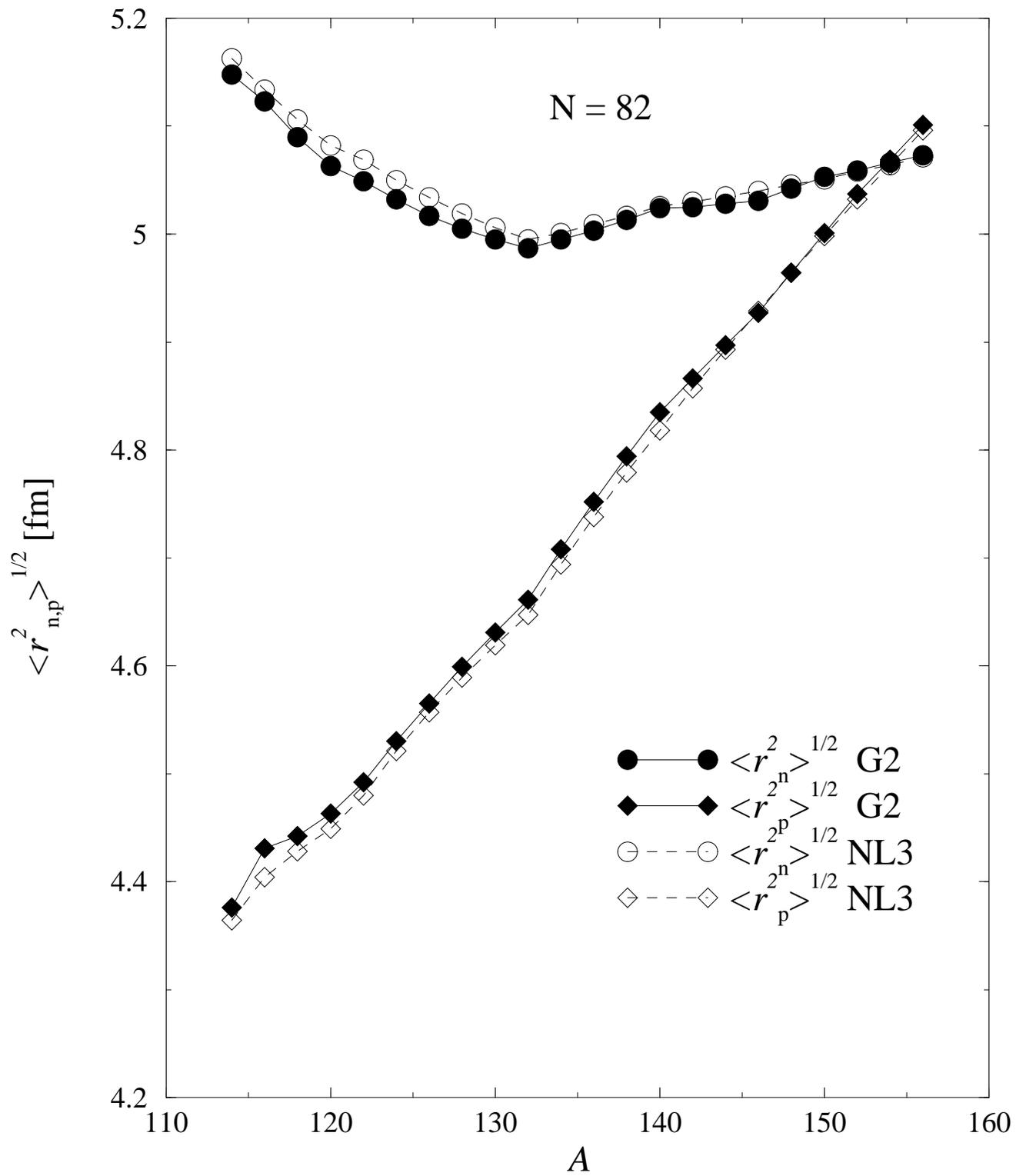
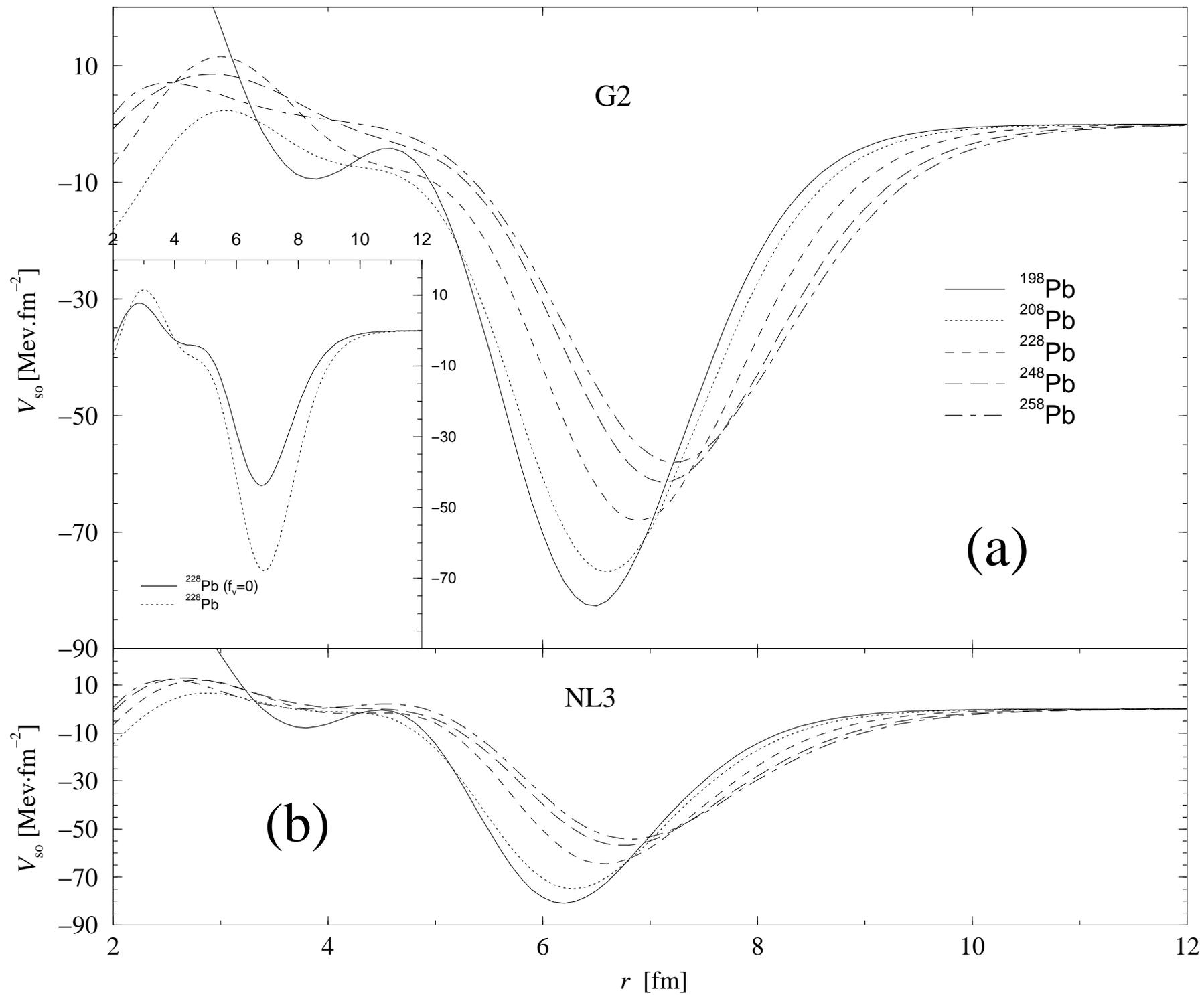
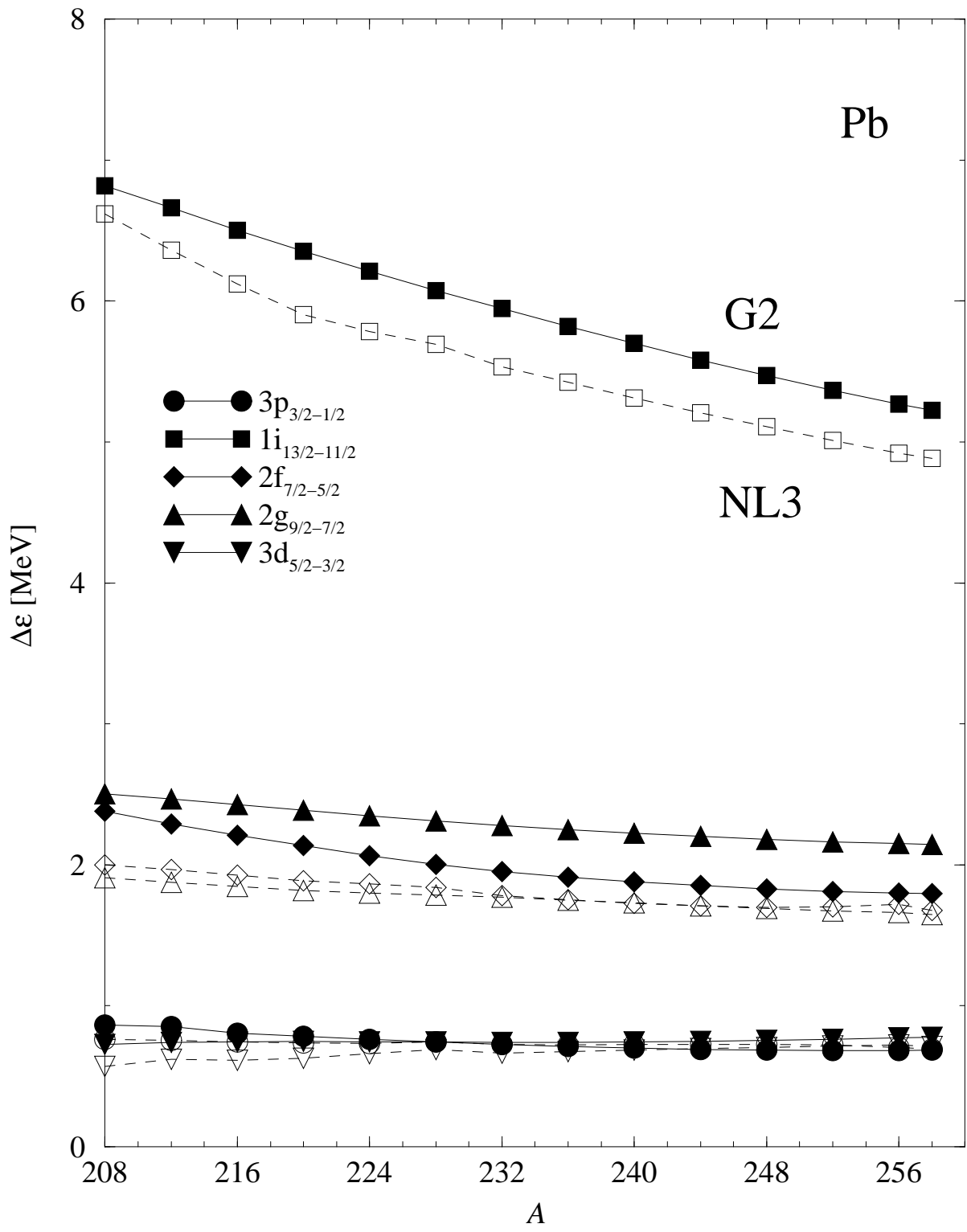
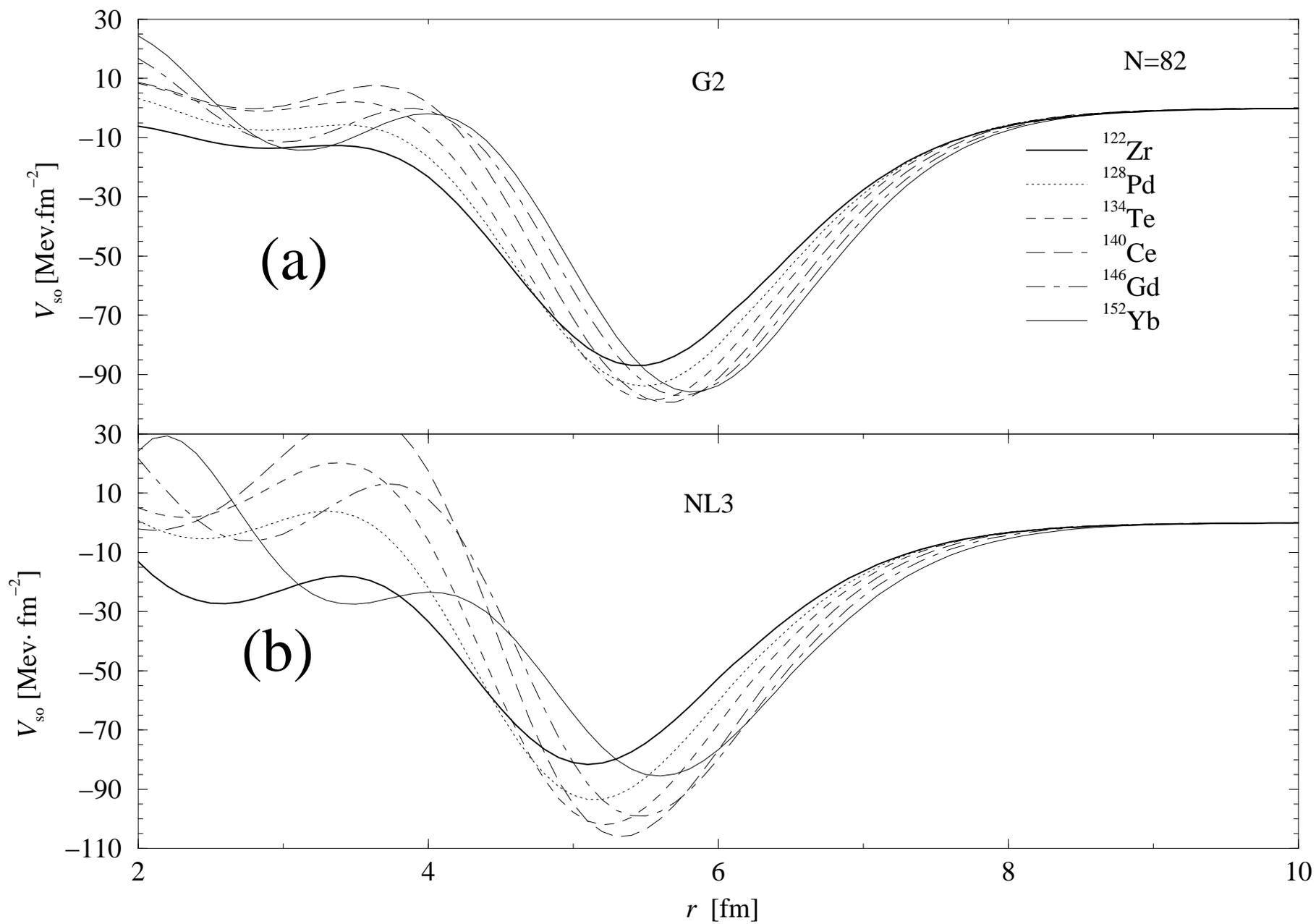
\end{document}